\documentclass[onecollarge]{svjour2}       
\smartqed  

\usepackage[dvipdfmx]{graphicx}

\usepackage{amsmath, amssymb}
\usepackage{amsfonts}
\usepackage{multicol}
\usepackage{braket}

\usepackage{ascmac}
\usepackage{fancybox}
\usepackage{comment}

\usepackage{cite}

\newcommand{\eq}[1]{\begin{equation} #1 \end{equation}}
\newcommand{\eqa}[2]{\begin{equation} #1 \label{#2} \end{equation}}
\newcommand{\balign}[1]{\begin{align} #1 \end{align}}

\newcommand{\fn}{\footnote}

\newcommand{\todayd}{\the\year/\the\month/\the\day}
\newcommand{\del}{\partial}
\newcommand{\bib}{\bibitem}
\newcommand{\appnum}[1]{\renewcommand{\theequation}{#1.\arabic{equation} }
\setcounter{equation}{0}}

\newcommand{\const}{\mathrm{const}}
\newcommand{\lmd}{\lambda}

\newcommand{\lb}{\label}
\newcommand{\nt}{\notag}
\newcommand{\Tr}{\mathrm{Tr}}

\newcommand{\eref}[1]{Eq.~\eqref{#1}}

\newcommand{\sref}[1]{Sec.~\ref{s:#1}}
\newcommand{\cref}[1]{Chap.~\ref{c:#1}}

\newcommand{\bpf}[1]{\begin{proof}: #1 \qed \end{proof}}

\newcommand{\bel}{\begin{easylist}}
\newcommand{\eel}{\end{easylist}}

\def \({\left(}
\def \){\right)}
\newcommand{\la}{\left\langle}
\newcommand{\ra}{\right\rangle}
\def \[{\left[}
\def \]{\right]}
\newcommand{\abs}[1]{\left|#1\right|}

\newcommand{\sumtwo}[2]%
{\mathop{\sum_{#1}}_{#2}}
\newcommand{\sumthree}[3]%
{\mathop{\mathop{\sum_{#1}}_{#2}}_{#3}}
\newcommand{\sumfour}[4]%
{\mathop{\mathop{\mathop{\sum_{#1}}_{#2}}_{#3}}_{#4}} 
\newcommand{\prodtwo}[2]%
{\mathop{\prod_{#1}}_{#2}}
\newcommand{\mintwo}[2]%
{\mathop{\min_{#1}}_{#2}}
\newcommand{\maxtwo}[2]%
{\mathop{\max_{#1}}_{#2}}
\newcommand{\maxthree}[3]%
{\mathop{\mathop{\max_{#1}}_{#2}}_{#3}}
\newcommand{\limtwo}[2]%
{\mathop{\lim_{#1}}_{#2}}
\newcommand{\suptwo}[2]%
{\mathop{\sup_{#1}}_{#2}}
\newcommand{\supthree}[3]%
{\mathop{\mathop{\sup_{#1}}_{#2}}_{#3}}
\newcommand{\supfour}[4]%
{\mathop{\mathop{\mathop{\sup_{#1}}_{#2}}_{#3}}_{#4}} 
\newcommand{\inftwo}[2]%
{\mathop{\inf_{#1}}_{#2}}
\newcommand{\infthree}[3]%
{\mathop{\mathop{\inf_{#1}}_{#2}}_{#3}}
\newcommand{\inffour}[4]%
{\mathop{\mathop{\mathop{\inf_{#1}}_{#2}}_{#3}}_{#4}} 

\newcommand{\bsp}{\boldsymbol{p}}

\newcommand{\bsx}{\boldsymbol{x}}

\newcommand{\bsB}{\boldsymbol{B}}

\newcommand{\ep}{\varepsilon}

\newcommand{\Di}{\mathit{\Delta}}

\newcommand{\tlr}{\tilde{R}}

\newcommand{\hsgm}{\hat{\sigma}}

\newcommand{\dsgm}{\dot{\sigma}}
\newcommand{\QH}{Q_{\rm H}}
\newcommand{\QL}{Q_{\rm L}}
\newcommand{\bL}{\beta_{\rm L}}
\newcommand{\bH}{\beta_{\rm H}}
\newcommand{\etac}{\eta_{\rm C}}
\newcommand{\Jq}{J^{\rm q}}
\newcommand{\Jn}{J^{\rm n}}

\newcommand{\Gd}{\Gamma^\dagger }

\def\rnum#1{\resizebox{0.5em}{\height}{\expandafter{\romannumeral #1}}}
\def\Rnum#1{\resizebox{0.5em}{\height}{\uppercase\expandafter{\romannumeral #1}}}

\begin{document}

\title{Fundamental relation between entropy production and heat current}


\author{Naoto Shiraishi and Keiji Saito}


\institute{Naoto Shiraishi and Keiji Saito \at
              Department of Physics, Keio University, 3-14-1 Hiyoshi, Yokohama 223-8522, Japan \\
             \email{shiraishi@rk.phys.keio.ac.jp}           
          }

\date{Received: date / Accepted: date}

\maketitle

\begin{abstract}

We investigate the fundamental relation between entropy production rate and the speed of energy exchange between a system and baths in classical Markov processes.
We establish the fact that quick energy exchange inevitably induces large entropy production in a quantitative form.
More specifically, we prove two inequalities on instantaneous quantities: 
One is applicable to general Markov processes induced by heat baths, and the other is applicable only to systems with the local detailed-balance condition but is stronger than the former one.
We demonstrate the physical meaning of our result by applying to some specific setups.
In particular, we show that our inequalities are tight in the linear response regime.

\end{abstract}

\section{Introduction}\label{intro}

Entropy production is a key quantity in nonequilibrium statistical mechanics, which characterizes the degree of irreversibility of thermodynamic processes.
In the last two decades, much effort has been devoted in the field of stochastic thermodynamics to clarify the deep connection between the entropy production and the path probability, which is clearly manifested by the celebrated fluctuation theorem~\cite{ECM93, Kur98, Jar00}.
Controlling entropy production is also intriguing and practically important in heat-related devices including heat engines and thermoelectric generators~\cite{MS96, MSS97, Maj04, ST08, CMP08, BSC11, Shi15, TH17, Shi17, BCSW}.

Although it is well known that slowing the speed of operation suppresses dissipation, it has not been clarified whether slowing the speed of operation is the {\it only} way to suppress dissipation.
Conventional thermodynamics gives no information on the speed of operation.
In the linear irreversible thermodynamics, even the coexistence of zero dissipation and finite power is not formally excluded if the time-reversal symmetry is broken~\cite{BSC11} (see also \ref{s:linear}).
After Ref.~\cite{BSC11}, some researchers investigate the idea to realize a heat engine with the Carnot efficiency at finite power~\cite{Alla13, CF16, Pon16, PE17, Joh17}.
In contrast, the existence of a trade-off relation between the speed of operation and the amount of dissipation has been strongly suggested by many specific models and setups, including endoreversible thermodynamics~\cite{CA75, And84}, thermoelectric transport with a magnetic field~\cite{SB12, BSS13, BS13, BBC13, BraS15, Yam16}, systems under time-asymmetric periodic driving~\cite{BSS15, PB15, PCB16, PS17}, overdamped Langevin systems~\cite{SS97, Aur12, Raz16}, and the thermodynamic uncertainty relation for stationary Markov jump systems~\cite{BarS15, Ging16, GRH17, PRS17, HG17, DS17, DS18}.
The first three classes of studies~\cite{CA75, And84, SB12, BSS13, BS13, BBC13, BraS15, Yam16, BSS15, PB15, PCB16, PS17} are restricted to the linear response regime, and the latter two classes of studies~\cite{SS97, Aur12, Raz16, BarS15, Ging16, GRH17, PRS17, HG17, DS17, DS18} require time-reversal symmetry.
Taking into account these backgrounds, a clear general picture on the speed of thermodynamic transformation is obviously necessary for further understanding of nonequilibrium thermodynamics.

In this paper, in line with our preceding letter~\cite{SST16}, we clarify the general principle of trade-off between speed and dissipation.
We derive universal inequalities on entropy production and heat current, which manifest the fact that quick energy exchange between the system and a bath inevitably accompanies much dissipation.
Our result is applicable to, for example, systems with broken time-reversal symmetry, systems beyond the linear response regime, systems under transient and time-dependent operations, as long as the system is described by classical or quantum Markov processes.
We also demonstrate the physical meaning of our inequalities in Langevin systems and systems in the linear response regime.
An important application of our result is to heat engines, in which we derive a universal trade-off inequality between power and efficiency.
Our findings solve in negative the problem whether a finite power engine attain the Carnot efficiency.

\

This paper is organized as follows.
\sref{framework} is devoted to an introductory review of Markov jump processes and stochastic thermodynamics.
Some symbols and key quantities are also introduced in this section.
In \sref{Markov-set}, we describe our setup and two main inequalities between entropy production and heat current:
One is general and the other is for systems with the local detailed-balance condition, while the latter is stronger than the former one.
We first demonstrate them in a simple setup, a single stochastic particle with a single bath, and then we move to a general setup.
We prove these inequalities in \sref{Markov-result}.
We first show the proof for the simple case in detail, and then show how this proof is generalized to the general case.
These two sections serve as a pedagogical rederivation of the results of Ref.~\cite{SST16}.

In \sref{theta}, we clarify its physical meaning of the coefficient $\Theta$ in some specific models.
Subsequent three sections are devoted to some applications and extensions of our inequality.
In \sref{Markov-eff}, we discuss the application of the inequalities to heat engines, which yields a universal trade-off relation between efficiency and power of heat engines.
We also discuss unusual behavior of power near the Carnot efficiency, which is sometimes confused as the coexistence of finite power and the Carnot efficiency.
In \sref{Markov-genform}, we generalize our inequalities.
The generalized inequalities concern entropy production and a time derivative of any quantity.
In \sref{quantum}, we extend our results to quantum Markov processes described by the Lindblad equation.

\section{Brief review of stochastic thermodynamics}\lb{s:framework}

\subsection{Framework of Markov process}
\subsubsection{Master equation}

Throughout this paper, we consider a classical Markovian system attached to some heat baths except otherwise noted.
A process is called Markovian if the time evolution of probability distribution of states depends only on the present probability distribution, not on its history.
The Markov process describes a variety of phenomena including Brownian particles, molecular motors, gas in a cylinder with thermal walls, and quantum dots in the classical regime.
From the perspective of physics, the Markov property means quick equilibration of heat baths.

The dynamics of a classical Markovian system is known to be well described by a Markov jump process with discrete states $\{ w\}$.
If the system is described with the continuous space (e.g., Langevin systems), we first take a proper discretization and then take the continuum limit.
This procedure works for both stochastic dynamics and deterministic Hamiltonian dynamics, which we shall briefly explain in the next subsection and discuss in detail in \ref{s:discre}.
Thus, we safely restrict our setup to the case with discrete states without loss of generality.

Let $p_{w,t}$ be the probability distribution of the state $w$ at time $t$.
The time evolution of the probability distribution $p_{w,t}$ is given by the following master equation
\eqa{
\frac{d}{dt}p_{w,t}=\sum_{w'}R_{ww'}p_{w',t},
}{master}
where $R_{ww'}$ represents the transition matrix.
The off-diagonal elements of a transition matrix satisfy {\it nonnegativity}; $R_{ww'}\geq 0$.
The term $R_{ww'}$ represents the conditional probability of jump from the state $w'$ to $w$ per unit time under the condition that the present state is $w'$.
Hence, the probability of the jump $w'\to w$ per unit time is given by $R_{ww'}p_{w',t}$.
The diagonal elements of a transition matrix $R_{w'w'}:=-\sum_{w(\neq w')}R_{ww'}<0$ represent the escape rate from the state $w'$ to another state per unit time.
We refer to the condition 
\eqa{
\sum_w R_{ww'}=0
}{normalization}
as the {\it normalization condition}.
A matrix is a transition matrix if it satisfies the nonnegativity and the normalization condition.
The transition matrix can be time-dependent in general, while we sometimes omit time-dependence in $R$ unless necessary.

Notably, the transition matrix is a linear operator, and stochastic driving from multiple baths is described by summation of transition matrices with each bath.
For example, if a system is attached to two heat baths 1 and 2 with two transition matrices $R^1$ and $R^2$, then the time evolution of the system is given by
\eq{
\frac{d}{dt}p_{w,t}=\sum_{w'}(R^1_{ww'}+R^2_{ww'})p_{w',t}.
}
Such decomposition also works for particles in many-particle systems.
If a system consists of $M$ particles and the stochastic process of the $i$-th particle is described by $R^i$, then the transition matrix of the whole system is given by $R=\sum_i R^i$.

\subsubsection{Discretization and continuum limit}\lb{s:discre-main}

In case of systems in continuous space, we take proper discretization of position and momentum space such that its continuum limit recovers the original dynamics of probability distribution.
Since the procedure of discretization and continuum limit is slightly technical, we here only show the corresponding transition rates in the discrete space and leave the details in \ref{s:discre}.

We here show the discretization procedure for a general Markov process of a single particle in one-dimensional continuous space, which is known to be described by the Kramers equation.
The extension to the case with multi-particle in higher dimension is straightforward.
The time-evolution of the probability distribution is given by
\eqa{
\frac{d}{dt}P( x ,  p)= \[ -\frac{ p}{m} \cdot \frac{\del}{\del  x} +\frac{\del}{\del  p} \cdot \( \frac{\gamma  p}{m} - F(x,p) \) +\frac{\gamma}{\beta }\frac{\del^2}{\del { p}^2}  \] P( x ,  p),
}{FP-main}
where $x$ and $ p$ are the position and momentum of the particle, and $\gamma$, $\beta$, $m$ are the friction coefficient, the inverse temperature, and the mass of the particle, respectively.
$F(x,p)$ represents the force acting on the particle, which includes both external and internal force.

We decompose the right-hand side of \eref{FP-main} into the Hamiltonian part
\eq{
\[ -\frac{ p}{m} \cdot \frac{\del}{\del  x} - \frac{\del}{\del  p} \cdot F(x,p) \] P(x,p)
}
and the dissipative part
\eq{
\[ \frac{\del}{\del  p} \cdot \frac{\gamma  p}{m} +\frac{\gamma}{\beta }\frac{\del^2}{\del { p}^2}\] P(x,p).
}
The former describes the Hamiltonian equation:
\balign{
\frac{d}{dt}p=&F(x,p) \\
\frac{d}{dt}x=&\frac{p}{m},
}
and the latter stands for the dissipative dynamics.

We now write down the corresponding transition matrices.
We first treat the deterministic Hamiltonian dynamics.
We discretize the $p-x$ phase space by the $\ep \times \ep '$ lattice.
A single state is determined by a pair of position and momentum, $(x,p)$.
Supposing $p>0$ and $F(x,p)>0$, we set the transition matrix of $(x,p)$ as
\balign{
R_{(x,p+\ep),(x,p)}&:=\frac{1}{\ep}F(x,p), \lb{disc-det1-main} \\
R_{(x+\ep',p),(x,p)}&:=\frac{1}{\ep'}\frac{p}{m}.\lb{disc-det2-main}
}
We remark that the inverse transitions do not occur (i.e., $R_{ (x,p),(x,p+\ep)}=0$ and $R_{(x,p),(x+\ep',p)}=0$).
The discretization of the dissipative part is given in a similar manner.
The transition matrix from a state with momentum $p$ to $p\pm \ep$ is given by
\eqa{
{R}_{p\pm \ep ,p}=\frac{\gamma}{\beta \ep^2}e^{-\frac{\beta}{4m}((p\pm \ep )^2-p^2)}=\frac{\gamma}{\beta \ep^2} e^{O(\ep)}.
}{disc-rate-main}
It is straightforward to recover the Kramers equation by taking the continuum limit $\ep\to 0$.


In summary, both dissipative and Hamiltonian dynamics can be well described by Markov jump processes with discrete states.
Notably, the dynamics given by Eqs.~\eqref{disc-det1-main} and \eqref{disc-det2-main} is stochastic with finite $\ep$ and $\ep'$, the dynamics becomes deterministic Hamiltonian dynamics in the continuum limit $\ep, \ep'\to 0$.
This is possible because the fluctuation due to the stochasticity converges to zero in the continuum limit.
Intuitively speaking, spatial space is divided more and more finely, and accordingly the number of jumps increases with keeping its average displacement, which results in vanishing fluctuation of position due to the law of large numbers.

The general form of the master equation for a $M$-particle system attached to $k$ baths including the Hamiltonian dynamics reads 
\eqa{
\frac{d}{dt}p_{w,t}=\sum_{w'} \( R_{ww'}^{0, \lmd (t)}+\sum_{\nu=1}^k \sum_{i=1}^M R_{ww'}^{\nu, i, \lmd (t)}\) p_{w',t},
}{R-decomp}
where $R_{ww'}^{0, \lmd (t)}$ corresponds to deterministic Hamiltonian dynamics, and $R_{ww'}^{\nu, i, \lmd (t)}$ represents the stochastic dynamics of the $i$-th particle induced by the $\nu$-th bath.
The parameter $\lmd (t)$ represents a control parameter.
It is noteworthy that $R^{0, \lmd (t)}$ acts on all particles simultaneously, while the transition matrix corresponds to the dissipative part can be decomposed into that of a single particle $R^{\nu, i, \lmd (t)}$ .
We sometimes write $R^\mu$ with $\mu=(\nu,i)$ or $\mu=0$.

\subsection{Framework of stochastic thermodynamics}

\subsubsection{Shannon entropy}

We now introduce some thermodynamic quantities in stochastic Markov processes.
We first define the entropy of the system following the formalism of stochastic thermodynamics.
In stochastic processes, the state of the system takes the form of a probability distribution $p$ on possible states $\{ w\}$.
We define the entropy of the system by the Shannon entropy:
\eq{
H(p):=-\sum_w p_w \ln p_w.
}

The Shannon entropy is first introduced in the information theory~\cite{CTbook}, which measures the degree of uncertainty of events, or states of a system.
If the system always takes a single state (i.e., $p_w$ takes one for a particular state $w$, and takes zero for all other states), the Shannon entropy takes the minimum value, zero.
In contrast, if the probability distribution is maximally mixed on possible $N$ states (i.e., $p_w=1/N$ for all states), the Shannon entropy takes the maximum value, $\ln N$.
In addition, if the probability distribution is the canonical distribution, then the Shannon entropy is equal to the conventional thermodynamic entropy.
Furthermore, as we will see later, by defining entropy of the system by the Shannon entropy, we can obtain the second law of thermodynamics.
On the basis of the above facts, we employ the Shannon entropy as the entropy of the system in stochastic thermodynamics.

\subsubsection{Heat and work}

We next define the heat and work in stochastic thermodynamics such that they satisfy the first law of thermodynamics.
In these definitions, the energy change in the system is decomposed into that caused by jumps and that caused by the change in the control parameter, which correspond to the heat and work respectively.

We first define the heat in stochastic thermodynamics.
If a heat bath induces a transition $w'\to w$, then the heat absorbed by the heat bath is simply defined as $E_{w'}-E_{w}$, where $E_w$ is the energy of the state $w$.
The heat current from the system to the $\nu$-th bath is written as
\eqa{
\Jq_\nu:=\sum_{w\neq w'}(E_{w'}-E_w)R^\nu_{ww'}p_{w'}=-\sum_{w,w'}E_{w}R^\nu_{ww'}p_{w'}.
}{def-jqnu}
In the second equality, we used the normalization condition $\sum_{w}R^\nu_{ww'}=0$.
If the transition matrix is decomposed into each particle, the heat current with the $i$-the particle to the $\nu$-th bath is given by 
\eqa{
\Jq _{\nu,i}:=-\sum_{w,w'}E_{w}R^{\nu,i}_{ww'}p_{w'},
}{def-jqmu}
whose sum over all particles $i$ yields the total heat current to the $\nu$-th bath: $\Jq _{\nu}=\sum_{i=1}^M \Jq _{\nu,i}$.
In this paper, we also refer to $\Jq _{\nu,i}$ as $\Jq _\mu$ for the sake of notational simplicity.

We next define the extracted work in stochastic thermodynamics.
The energy of the state $w$ is in general time-dependent through the change of the control parameter $\lmd (t)$.
To manifest this fact, we explicitly write the $\lmd$-dependence of the energy as $E_w^\lmd$.
The work extraction per unit time is defined as the change in energy through the change in the control parameter:
\eq{
\dot{W}:=-\sum_w p_w \frac{dE_w^\lmd}{d\lmd}\frac{d\lmd}{dt}.
}
With these definitions, the first law of thermodynamics is indeed satisfied:
\eq{
\frac{d}{dt}\la E\ra =-\sum_\nu \Jq_\nu-\dot{W},
}
where $\la E\ra :=\sum_w E_wp_w$ is the average of energy.
The left-hand side means the change in the energy of the system.

\subsubsection{Requirement for transition matrix}\lb{s:req-matrix}

Throughout this paper we require the invariance of the canonical distribution for each bath and for each particle.
This requirement reflects the fact that a heat bath does not change the state of a system in equilibrium with the same temperature.
By denoting the inverse temperature of the $\nu$-th  bath by $\beta_\nu$, the above condition for the $i$-th component and the $\nu$-th bath reads
\eqa{
\sum_{w'} R^{\nu, i,\lmd (t)}_{ww'}e^{-\beta_\nu E_{w'}^{\lmd (t)}}=0.
}{cano-inv-sto}
In case of a particle bath, the canonical distribution is replaced by the grand canonical distribution.
The transition matrix corresponding to the Hamiltonian dynamics, $R^{0, \lmd (t)}$, is required to keep the uniform distribution invariant:
\eqa{
\sum_{w'} R^{0,\lmd (t)}_{ww'}=0.
}{cano-inv-det}
This condition reflects the fact that the Hamiltonian dynamics keeps the energy shell invariant.
We remark that we have not required that the canonical distribution or the uniform distribution is the unique stationary (invariant) distribution.

\

In some cases, we impose a stronger requirement: the local detailed-balance condition
\eqa{
R^{\nu, i,\lmd (t)}_{ww'}e^{-\beta_\nu E_{w'}^{\lmd (t)}}=R^{\nu, i, \lmd (t)}_{w'w}e^{-\beta_\nu E_{w}^{\lmd (t)}}
}{db}
for any $\nu$, $i$, $w$ and $w'$.
The local detailed-balance condition means that in the canonical distribution no probability current exists between any pair of two states $w$ and $w'$.
Note that the Hamiltonian part $R^0$ does not satisfy the local detailed-balance condition.

We emphasize that we do {\it not} take time-reversal of the states and the transition rates.
Thus, the local detailed-balance condition \eqref{db} is in general violated in systems with parity-odd variables (e.g., momentum) or parity-odd fields (e.g., a magnetic field).
Due to this, the condition \eqref{db} is sometimes called {\it time reversal symmetry}.

\subsubsection{Entropy production and second law of thermodynamics}\lb{s:second-law}

We now introduce the entropy production rate, which is one of the most important quantities in stochastic thermodynamics.
The entropy production rate is defined as
\eqa{
\dot{\sigma}:=\frac{dH(p_t)}{dt}+\sum_{\nu=1}^k \beta_\nu \Jq _\nu .
}{def-sigma}
The first term is the entropy increase of the system, and the second term is sum of the entropy increase of all baths.
Hence, the entropy production rate can be regarded as the rate of entropy increase of the composite system of the system and the baths.

We confirm that the entropy production rate is nonnegative, which is the second law of thermodynamics in stochastic thermodynamics.
To demonstrate this, we introduce the {\it dual transition matrix} for each transition matrix defined as
\eq{
\tlr^\mu _{w'w}:=e^{\beta _\nu (E_w-E_{w'})}R^\mu_{ww'}.
}
Due to the invariance of canonical distribution, the dual transition matrix satisfies the normalization condition:
\eq{
\sum_{w'}\tlr^\mu _{w'w}=e^{\beta _\nu E_w}\sum_{w'}R^\mu_{ww'}e^{-\beta_\nu E_{w'}}=0
}
for any $w$.
Since $\tlr^\mu$ obviously satisfies the nonnegativity, we confirm that the matrix $\tlr^\mu$ is indeed the transition matrix.
If the original transition matrix satisfies the local detailed-balance condition, the dual transition matrix reduces to the original transition matrix; $\tlr^\mu _{ww'} =R^\mu_{ww'}$.
Although this relation generally violated in systems without the local detailed-balance condition, the diagonal elements of the original and dual transition matrix are always the same by definition: 
\eq{
\tlr ^\mu_{ww}=R^\mu_{ww}
}
for any state $w$.
The dual transition matrix is an artificial but useful tool to prove some relations important in physics.

Using the dual transition matrix, the nonnegativity of the entropy production rate is proven as
\balign{
\dsgm =&-\sum_w \frac{d}{dt}(p_w \ln p_w)+\sum_{\mu} \beta_\nu \sum_{w\neq w'}(E_{w'}-E_w)R^\mu_{ww'}p_{w'} \nt \\
=&-\sum_\mu \sum_{w,w'} R^\mu_{ww'}p_{w'}\ln p_w +  \sum_\mu \sum_{w\neq w'} \( \ln \frac{R^\mu_{ww'}}{\tlr^\mu_{w'w}}\) R^\mu_{ww'}p_{w'} \nt \\
=&\sum_\mu \sum_{w\neq w'}R^\mu_{ww'}p_{w'} \ln \frac{R^\mu_{ww'}p_{w'}}{\tlr^\mu_{w'w}p_{w}} \lb{dsgm-rel}  \\
\geq &0.
}
In the third line, we used the normalization condition; $\sum_{w} R^\mu_{ww'}p_{w'}\ln p_{w'}=0$.
In the last line, we used the nonnegativity of relative entropy~\cite{CTbook}
\eq{
D(p||q):=\sum_i p_i \ln \frac{p_i}{q_i}\geq 0
}
for any two distribution $p$, $q$ such that $\sum_i p_i=\sum_i q_i$, and the following relation:
\eq{
\sum_{w\neq w'}R^\mu_{ww'}p_{w'}=-\sum_{w'}R^\mu_{w'w'}p_{w'}=-\sum_{w}\tlr^\mu_{ww}p_{w}=\sum_{w\neq w'}\tlr^\mu_{w'w}p_w.
}
The expression in the second line \eqref{dsgm-rel} is also useful, and we will use this form in the derivation of our main results.

\section{Setup and main result}\lb{s:Markov-set}

\subsection{Simple case: a single stochastic particle with single-bath}

To demonstrate our main results, we first consider a simple setup: a Markov process driven by a single heat bath.
We consider a general case in the next subsection.

Consider a Markov process with discrete states $\{ w\}$ driven by a single heat bath with inverse temperature $\beta$ without Hamiltonian dynamics.
Since the system has a single particle and attached to a single heat bath, the transition matrix $R$ in \eref{master} itself satisfies the invariance of the canonical distribution
\eqa{
\sum_{w'}R_{ww'}e^{-\beta E_{w'}}=0.
}{cano-inv-simple}
The heat current from the system to the bath \eqref{def-jqnu} and the entropy production rate \eqref{def-sigma} reads
\balign{
\Jq (t)&:=-\sum_{w,w'}E_{w}R_{ww'}p_{w',t}, \\
\dot{\sigma}(t)&:=\frac{d}{dt}H(p_t)+\sum \beta \Jq (t).
}

As proven in \sref{second-law}, the second law of thermodynamics claims that the entropy production rate is nonnegative: $\dsgm (t)\geq 0$.
On the other hand, we have a stronger inequality on the entropy production rate:

\

{\bf Theorem 1.1 (simple case)}.
Consider a Markov process where the canonical distribution is invariant (i.e., the condition \eqref{cano-inv-simple}).
Then, the heat current and the entropy production rate satisfy
\eqa{
|\Jq  (t)|\leq \sqrt{\Theta^{(1)} (t)\dsgm (t)}
}{Jsigma-1-simple}
with
\eqa{
\Theta^{(1)} (t):=\frac{1}{c_0} \sum_{w\neq w'}(\Di E_w)^2 (R_{ww'}p_{w',t}+R_{w'w}p_{w,t}).
}{theta1-def-simple}
Here, $\Di E_w:=E_w-\la E_w\ra$ represents the energy fluctuation from its ensemble average $\la E_w\ra =\sum_w E_w p_{w,t}$ and $c_0:=8/9$.
We remark that the coefficient $c_0$ is not the best one and the best coefficient is $c^*:=0.89612\cdots$.
This point is discussed in \ref{s:best-coeff}.

\

{\bf Theorem 1.2 (simple case)}.
Consider a Markov process with the local detailed-balance condition: $R_{ww'}e^{-\beta E_{w'}}=R_{w'w}e^{-\beta E_{w}}$ for any $w$ and $w'$.
Then, the heat current and the entropy production rate satisfy
\eqa{
|\Jq  (t)|\leq \sqrt{\Theta^{(2)} (t)\dsgm (t)}
}{Jsigma-2-simple}
with
\eqa{
\Theta^{(2)} (t):=\frac{1}{2} \sum_{w\neq w'}(E_w-E_{w'})^2 R_{ww'}p_{w',t}.
}{theta2-def-simple}

\

The physical meaning of the inequalities~\eqref{Jsigma-1-simple} and \eqref{Jsigma-2-simple} is clear.
If we exchange energy between a bath and a system quickly, then much dissipation (irreversible energy loss) must be generated.
In other words, this inequality manifests a trade-off between the speed of energy exchange and dissipation.

For short time duration $\Di t$, the first moment $\sum_{w\neq w'}(E_w-E_{w'}) R_{ww'}p_{w',t} \Di t$ is the average energy exchange in this duration, and the second moment is given by $\sum_{w\neq w'}(E_w-E_{w'})^2 R_{ww'}p_{w',t}\Di t$.
Hence, the coefficient $\Theta^{(2)}$ is the half of the second moment (rate) of the instantaneous energy exchange between the bath and the system.

We remark that if a system has parity-odd variables (e.g., momentum) or a parity-odd field (e.g., magnetic field), the local detailed-balance condition no longer holds in general and only Theorem 1.1 is satisfied.

\subsection{General case}

We now describe a general Markov process of a system of $M$ components (particles) induced by $k$ heat baths.
The state of the total system $w$ is a combination of the states of $M$ particles: $w=(w^1, w^2, \cdots, w^M)$.
If one is interested in only a small system, $M$ is set to 1.
The master equation is given in \eref{R-decomp}.
It is noteworthy that $R_{ww'}^{0, \lmd (t)}$ in \eref{R-decomp} does not contribute to the heat current because this dynamics is isolated and does not accompany heat baths.


To explain our main inequality, we introduce conditional probability distribution and conditional quantities.
First, we denote by $w^{-i}:=(w^1,\cdots ,w^{i-1}, w^{i+1}, \cdots ,w^M)$ the state of $w$ except the $i$-th particle.
Then, for a given probability distribution $p_w$ and a given particle $i$, the {\it conditional probability distribution of the $i$-th particle with respect to $w^{-i}$} is defined as
\eq{
p_{{w}^i |w^{-i}}:=\frac{p_{{w}^i, w^{-i}}}{p_{w^{-i}}}
}
with $p_{w^{-i}}:=\sum_{{w'}^i}p_{{w'}^i, w^{-i}}$.
Here, since the pair $(w^i, w^{-i})$ specifies the state of all the particle, $p_{w^i, w^{-i}}$ is the same as the probability distribution $p_w$.
Using this notation, the conditionalized average is defined as
\eqa{
\la A\ra_{w^{-i}}:=\sum_{{w'}^i} A({w'}^i,w^{-i})p_{{w'}^i |w^{-i}}.
}{cond-def}

\

{\bf Theorem 1.1}.
Consider a Markov process where the canonical distribution is invariant (i.e., the conditions \eqref{cano-inv-sto} and \eqref{cano-inv-det}).
Then, the heat current \eqref{def-jqnu} and the entropy production rate \eqref{def-sigma} satisfy
\eqa{
\sum_{\nu =1}^k |\Jq _\nu (t)|\leq \sqrt{\Theta^{(1)} (t)\dsgm (t)}
}{Jsigma-1}
with
\eqa{
\Theta^{(1)} (t):=\frac{1}{c_0}\sum_\mu \sum_{w\neq w'}(\Di E_w^{\mu, \lmd (t)})^2 (R_{ww'}^{\mu, \lmd (t)}p_{w',t}+R_{w'w}^{\mu, \lmd (t)}p_{w,t}).
}{theta1-def}
Here, $\Di E_w^{\mu, \lmd (t)}$ represents the energy fluctuation of the $i$-th particle under the conditional probability distribution defined as
\eqa{
\Di E_w^{\mu, \lmd (t)}:=E_w^{\lmd (t)}-\la E^{\lmd (t)}\ra_{t, w^{-i}},
}{def-DiE}
where the label of a particle $i$ is set to the same as that in $\mu =(i,\nu)$, and $\la \cdot \ra_{t, w^{-i}}$ is the same as \eref{cond-def} for the probability distribution at time $t$.
If $\mu =0$, we define $\Di E_w^{\mu, \lmd (t)}=0$.

\

{\bf Theorem 1.2}.
Consider a Markov process with the local detailed-balance condition \eqref{db}.
Then, the heat current and the entropy production rate satisfy
\eqa{
\sum_{\nu =1}^k |\Jq _\nu (t)|\leq \sqrt{\Theta^{(2)} (t)\dsgm (t)}
}{Jsigma-2}
with
\eqa{
\Theta^{(2)} (t):=\frac{1}{2}\sum_{w\neq w'}(E_w^{\lmd (t)}-E_{w'}^{\lmd (t)})^2 R_{ww'}^{\lmd (t)}p_{w',t}.
}{theta2-def}

\

The former inequality~\eqref{Jsigma-1} is applicable to any physical stochastic processes, while the latter one~\eqref{Jsigma-2} is applicable only to systems with the local detailed-balance condition.
By contrast, the latter inequality~\eqref{Jsigma-2} is stronger than the former one~\eqref{Jsigma-1}.

We remark that both $\Theta^{(1)}$ and $\Theta^{(2)}$ are shown to be finite under some physically plausible assumptions.
In addition, in the thermodynamic limit both $\Theta^{(1)}$ and $\Theta^{(2)}$ increase linearly with respect to the entropy production $\dsgm$ and the heat current $\Jq$, which means that our inequalities are still nontrivial relations in the macroscopic systems.
These facts are shown in \sref{fin-theta}.

\section{Proofs}\lb{s:Markov-result}

\subsection{Simple case: a single stochastic particle with single-bath}

The essence of the proofs can be seen in that of the simple case.
Therefore, we explain the proofs for the simple case in detail.
In this and next subsection, we drop the dependence of time $t$ and the control parameter $\lmd(t)$.

\subsubsection{Proof of Theorem 1.1 for the simple case}\lb{s:Markov-gen}

To prove \eref{Jsigma-1-simple}, we introduce a useful lemma.
The proof is given in \ref{s:lemma1}

\

{\bf Lemma 1}:
For two distributions $p_i$ and $q_i$ satisfying $\sum_i p_i=\sum_i q_i$, the (extended) relative entropy $D(p||q):=\sum_i p_i \ln (p_i/q_i)$ is bounded as
\eqa{
D(p||q)\geq c_0 \sum_i \frac{(p_i-q_i)^2}{p_i+q_i}
}{rel-ent-lemma}
with $c_0=8/9$.
The right-hand side of \eqref{rel-ent-lemma} without $c_0$ is named {\it triangular discrimination} in information theory~\cite{Tan05}.

\

We remark that $c_0=8/9$ is not tightest.
The best coefficient $c^*=0.896\cdots$ is obtained only numerically, which is discussed in \ref{s:best-coeff}.

We now prove the inequality~\eqref{Jsigma-1-simple}.

\bpf{
Using the Lemma 1, the entropy production rate \eqref{dsgm-rel} is evaluated as
\balign{
\dsgm 
=\sum_{w, w'} R_{ww'}p_{w'}\ln \frac{R_{ww'}p_{w'}}{\tlr_{w'w}p_{w}}
\geq& c_0 \sum_{w, w'}\frac{(R_{ww'}p_{w'}-\tlr_{w'w}p_{w})^2}{R_{ww'}p_{w'}+\tlr_{w'w}p_{w}}
=c_0 \sum_{w\neq w'}\frac{(R_{ww'}p_{w'}-\tlr_{w'w}p_{w})^2}{R_{ww'}p_{w'}+\tlr_{w'w}p_{w}}. \lb{Jsigma1-mid2}
}
In the last equality, we used $R_{ww}=\tlr _{ww}$.
The heat current $\Jq$ is transformed into
\balign{
\Jq 
&:=-\sum_{w,w'}E_w R_{ww'} p_{w'} \nt \\
&=-\sum_{w,w'}E_w (R_{ww'}p_{w'}-\tlr_{w'w}p_{w}) \nt \\
&=-\sum_{w,w'}\Di E_w (R_{ww'}p_{w'}-\tlr_{w'w}p_{w}) \nt \\
&=-\sum_{w\neq w'}\Di E_w (R_{ww'}p_{w'}-\tlr_{w'w}p_{w}), \lb{Jq-gen}
}
where we used the normalization condition ($\sum_w R_{ww'}=0$ and $\sum_{w'}\tlr_{w'w}=0$) in the second and third lines and $R_{ww}=\tlr _{ww}$ in the fourth line.

We then have the desired inequality:
\balign{
|\Jq |^2
&=\abs{\sum_{w\neq w'}\Di E_w (R_{ww'}p_{w'}-\tlr_{w'w}p_{w})}^2 \nt \\
&=\abs{\sum_{w\neq w'}\Di E_w \sqrt{R_{ww'}p_{w'}+\tlr_{w'w}p_{w}} \cdot \frac{R_{ww'}p_{w'}-\tlr_{w'w}p_{w}}{\sqrt{R_{ww'}p_{w'}+\tlr_{w'w}p_{w}}} }^2 \nt \\
&\leq \( \sum_{w\neq w'} (\Di E_w)^2 (R_{ww'}p_{w'}+\tlr_{w'w}p_{w}) \) \( \sum_{w\neq w'}\frac{(R_{ww'}p_{w'}-\tlr_{w'w}p_{w})^2 }{R_{ww'}p_{w'}+\tlr_{w'w}p_{w}} \) \nt \\
&\leq \( \sum_{w\neq w'} (\Di E_w)^2 (R_{ww'}p_{w'}+\tlr_{w'w}p_{w}) \) \frac{1}{c_0} \dsgm \nt \\
&= \( \sum_{w\neq w'} (\Di E_w)^2 (R_{ww'}p_{w'}+R_{w'w}p_{w}) \) \frac{1}{c_0} \dsgm \nt \\
&=\Theta^{(1)} \dsgm . \lb{Jsigma1-mid}
}
Here, we used the Schwarz inequality in the third line, \eref{Jsigma1-mid2} in the fourth line, and a relation suggested by the normalization condition 
\eq{
\sum_{w' (\neq w)} (\Di E_w)^2\tlr_{w'w}p_{w}=-(\Di E_w)^2\tlr_{ww}p_{w}=-(\Di E_w)^2R_{ww}p_{w}=\sum_{w' (\neq w)} (\Di E_w)^2R_{w'w}p_{w}
}
in the fifth line.
}

The key transformation of the entropy production rate is seen in the second equality of \eref{lemma-pep} in the derivation of Lemma 1.
Although a single summand of the relative entropy $p_i\ln p_i/q_i$ can be both positive and negative, the transformed summand $p_i\ln p_i/q_i+q_i-p_i$ is always nonnegative.
This allows us to evaluate relative entropy in a quadratic form.
In fact, the form of the summand $p_i\ln p_i/q_i+q_i-p_i$ is exactly the same as the ensemble average of the partial entropy production~\cite{SS15, SIKS15, SMS16, SS16} of a single transition.

\subsubsection{Proof of Theorem 1.2 in the simple case}\lb{s:Markov-db}

To prove the inequality~\eqref{Jsigma-2-simple}, we use another simple mathematical inequality.
The proof is given in \ref{s:lemma2}

\

{\bf Lemma 2}:
For $a,b>0$, we have an inequality
\eqa{
(a-b)\ln \frac{a}{b}\geq \frac{2(a-b)^2}{a+b}.
}{simple-lemma}

\

We now derive \eref{Jsigma-2-simple} under the local detailed-balance condition.

\bpf{
Using the Lemma 2, the entropy production rate is calculated as
\balign{
\dsgm  
=\sum_{w\neq w'} R_{ww'}p_{w'}\ln \frac{R_{ww'}p_{w'}}{R_{w'w}p_{w}}
=&\frac{1}{2}\sum_{w\neq w'} (R_{ww'}p_{w'}-R_{w'w} p_w)\ln \frac{R_{ww'}p_{w'}}{R_{w'w}p_{w}} \nt \\
\geq& \sum_{w\neq w'}\frac{(R_{ww'}p_{w'}-R_{w'w}p_{w})^2}{R_{ww'}p_{w'}+R_{w'w}p_{w}}.
}
The heat current $\Jq $ is also transformed into
\balign{
\Jq 
:&=-\sum_{w,w'}E_w R_{ww'} p_{w'} \nt \\
&=-\sum_{w,w'}E_w (R_{ww'}p_{w'}-R_{w'w}p_{w}) \nt \\
&=-\frac{1}{2}\sum_{w,w'} (E_w-E_{w'}) (R_{ww'}p_{w'}-R_{w'w}p_{w}) \nt \\
&=-\frac{1}{2}\sum_{w\neq w'}(E_w-E_{w'}) (R_{ww'}p_{w'}-R_{w'w}p_{w}). \lb{Jq-db}
}
In a similar manner to \eqref{Jsigma1-mid}, we obtain the desired inequality:
\balign{
|\Jq |^2
&=\abs{ \frac{1}{2} \sum_{w\neq w'} (E_w-E_{w'}) (R_{ww'}p_{w'}-R_{w'w}p_{w})}^2 \nt \\
&=\abs{\frac{1}{2} \sum_{w\neq w'} (E_w-E_{w'}) \sqrt{R_{ww'}p_{w'}+R_{w'w}p_{w}} \cdot \frac{R_{ww'}p_{w'}-R_{w'w}p_{w}}{\sqrt{R_{ww'}p_{w'}+R_{w'w}p_{w}}} }^2 \nt \\
&\leq \( \frac{1}{4}\sum_{w\neq w'} ( E_w-E_{w'})^2 (R_{ww'}p_{w'}+R_{w'w}p_{w}) \) \( \sum_{w\neq w'}\frac{(R_{ww'}p_{w'}-R_{w'w}p_{w})^2 }{R_{ww'}p_{w'}+R_{w'w}p_{w}} \) \nt \\
&\leq \Theta^{(2)} \dsgm .
}
}

The difference between with and without the local detailed-balance condition appears in the transformation of $\Jq$.
In \eref{Jq-gen} the heat current is written in terms of energy fluctuation $\Di E_w$, while in \eref{Jq-db} it is written in terms of energy difference $E_w-E_{w'}$.
This difference is crucial when we consider the continuum limit.

\subsection{General case}

We now consider the case of $M$ components with $k$ baths.
The inequalities for the general case, \eqref{Jsigma-1} and \eqref{Jsigma-2}, are respectively derived from \eqref{Jsigma-1-simple} and \eqref{Jsigma-2-simple} in a rather direct manner.

\bpf{
The entropy production rate $\dot{\sigma}:=dH(p_t)/dt+\sum_{\nu=1}^k \beta_\nu \Jq _\nu$ is decomposed into the contribution of the $i$-th particle and the $\nu$-th bath as
\eq{
\dsgm _\mu:=-\sum_{w,w'}R^{\mu}_{ww'}p_{w',t} \ln p_{w,t}+\beta_\nu \Jq _\nu,
}
which satisfies $\dsgm =\sum_\mu \dsgm _\mu$.
The result  \eqref{Jsigma-1-simple} or \eqref{Jsigma-2-simple} implies
\eq{
\abs{\Jq_{\mu}}\leq \sqrt{\Theta_\mu \dsgm_\mu},
}
where $\Theta_\mu$ takes $\Theta_\mu^{(1)}$ or $\Theta_\mu^{(2)}$ defined as
\balign{
\Theta_\mu^{(1)}&:=\frac{1}{c_0} \sum_{w\neq w'} (\Di E_w^{\mu})^2 (R_{ww'}^{\mu}p_{w',t}+R_{w'w}^{\mu}p_{w}), \lb{thetamu1}  \\
\Theta_\mu^{(2)}&:=\frac{1}{2} \sum_{w\neq w'} ( E_w-E_{w'})^2 R_{ww'}^{\mu}p_{w',t}.
}
We note $\sum_\mu \Theta_\mu^{(1)}=\Theta^{(1)}$ and $\sum_\mu \Theta_\mu^{(2)}=\Theta^{(2)}$.

Applying the Schwarz inequality, we arrive at the desired inequality:
\balign{
 \( \sum_{\nu} |\Jq _{\nu}| \) ^2
\leq \( \sum_{\mu} |\Jq _{\mu}| \) ^2
\leq \( \sum_{\mu} \sqrt{\Theta_{\mu}}\cdot \sqrt{\dsgm_{\mu}}\) ^2
\leq \( \sum_{\mu} {\Theta_{\mu}}\) \( \sum_{\mu} {\dsgm_{\mu}}\) 
&=  \Theta \dsgm .
}
}



\section{Concrete form of $\Theta$ in some specific systems}\lb{s:theta}

The coefficient $\Theta$ in the inequalities \eqref{Jsigma-1} and \eqref{Jsigma-2} are defined in a highly abstract way, and their physical interpretation has not yet been clarified.
In this section, we apply the obtained relation to some specific setups and clarify their physical meanings.

\subsection{Case of underdamped Langevin system}\lb{s:Lan-theta}

Consider an underdamped Langevin system with a magnetic field $\bsB$.
Corresponding Fokker-Planck equation reads
\eq{
\frac{d}{dt}P(\bsx , \bsp)=\sum_i \[ -\frac{\bsp^i}{m^i} \cdot \frac{\del}{\del \bsx^i} +\frac{\del}{\del \bsp^i} \cdot \( \frac{\gamma \bsp^i}{m^i} -F_i(\bsx, \bsp)+ \frac{1}{m^i} \bsB \times \bsp^i \) +\frac{\gamma}{\beta }\frac{\del^2}{\del {\bsp^i}^2}  \] P(\bsx , \bsp),
}
where $\bsx^i$ and $\bsp^i$ are the position and momentum of the $i$-th particle, $\bsx :=\{ \bsx^1,\bsx^2,\cdots \}$ and $\bsp :=\{ \bsp^1,\bsp^2,\cdots \}$ are the set of positions and momentums, $\gamma$, $\beta$, $m^i$ are the friction coefficient, the inverse temperature, and the mass of the $i$-th particle, respectively.
The force on the $i$-th particle $F_i(\bsx, \bsp)$ includes both the potential force $\del U(\bsx)/\del x_i$ and the external force.
We explicitly wrote down the Lorentzian force separately.
This is a general expression of dynamics of Markovian stochastic particles with a heat bath because the system size expansion always provides stochastic Markov processes in this form~\cite{Kambook}.

As seen in \sref{discre-main}, the Fokker-Planck operator can be decomposed into the Hamiltonian dynamics part
\eq{
-\frac{\bsp^i}{m^i} \cdot \frac{\del}{\del \bsx^i} +\frac{\del}{\del \bsp^i} \cdot \(  -F_i(\bsx, \bsp)+ \frac{1}{m^i} \bsB \times \bsp^i \) \nt
}
and the dissipative part
\eq{
\frac{\del}{\del \bsp^i} \cdot \frac{\gamma \bsp^i}{m^i} +\frac{\gamma}{\beta }\frac{\del^2}{\del {\bsp^i}^2}. \nt
}
We decompose the entropy production rate and heat current into the contributions from these two parts: $\dsgm=\dsgm_{\rm Ham}+\dsgm_{\rm dis}$ and $\Jq=\Jq_{\rm Ham}+\Jq_{\rm dis}$.
Due to the second law of thermodynamics, the Hamiltonian dynamics part should yield nonnegative entropy production rate: $\dsgm_{\rm Ham}\geq 0$.
In addition, the Hamiltonian part does not contribute to the heat current: $\Jq_{\rm Ham}=0$.
Hence, the inequality $\abs{\Jq_{\rm dis}}\leq \sqrt{\Theta_{\rm dis} \dsgm_{\rm dis}}$ directly implies $\abs{\Jq}\leq \sqrt{\Theta \dsgm}$ with $\Theta=\Theta_{\rm dis}$.
In other words, it suffices to show the inequality in the system where the time-evolution operator consists of only the dissipative part.

In the following, we treat only the dissipative part, and for simplicity we consider the case of one-dimensional single particle system with a single bath.
This simplification is justified because of the absence of interaction with other particles.
We use the discretized transition rate shown in \sref{discre-main}.
The transition rate from a state with momentum $p$ to $p\pm \ep$ is given by
\eq{
{R}_{p\pm \ep ,p}=\frac{\gamma}{\beta \ep^2}e^{-\frac{\beta}{4m}((p\pm \ep )^2-p^2)}=\frac{\gamma}{\beta \ep^2} e^{O(\ep)},
}
which is the same as \eref{disc-rate-main}.
Since this transition rate satisfies the local detailed-balance condition~\eqref{db}, the stronger inequality \eqref{Jsigma-2} is applicable to the underdamped Langevin systems.
We now calculate the energy fluctuation.
Using the expression
\eq{
(E_p-E_{p\pm \ep})^2=\( \frac{p^2}{2m}- \frac{(p\pm \ep )^2}{2m}\) ^2= \frac{(p\ep )^2}{m^2} +O(\ep ^3),
}
we have
\eq{
\frac{1}{2}\sum_\pm (E_p-E_{p\pm \ep})^2 {R}_{p\pm \ep ,p}=\frac{(p\ep )^2}{m^2}\frac{\gamma}{\beta \ep^2}+O(\ep)=\frac{\gamma p^2}{\beta m^2}+O(\ep ).
}
The continuum limit of the above relation yields an explicit expression of $\Theta^{(2)}$ as
\eqa{
\Theta^{(2)}= \la \frac{\gamma |\bsp|^2}{\beta m^2}\ra =\frac{2\gamma}{\beta m} \langle \hat{K} \rangle,
}{theta-underL}
where $\hat{K}$ represents the kinetic energy of the system.

\subsection{Case of overdamped Langevin systems}

We remark that the overdamped limit of $\Theta^{(2)}$ in \eref{theta-underL} diverges and \eref{Jsigma-2} does not provide meaningful information more than the second law.
To avoid this, we directly discretize the overdamped Langevin system.
For simplicity, we again consider the case of a single particle in one-dimensional space.
The time evolution of this system is described by
\eq{
\frac{d}{dt}P(x)=-\frac{1}{\gamma}\frac{\del}{\del x}\( F(x)P(x)\) +\frac{1}{\gamma \beta}\frac{\del^2}{\del x^2}P(x).
}
The discretized transition rate from $x$ to $x\pm \ep$ and the energy difference are given by
\balign{
R_{x\pm \ep ,x}=&\frac{\gamma}{\beta \ep^2} e^{O(\ep)}, \\
(E_x-E_{x\pm \ep})^2=&(F(x)\ep)^2+O(\ep^3).
}
Using this, we can calculate $\Theta^{(2)}$ in a similar manner and obtain an explicit expression of $\Theta^{(2)}$ as
\eq{
\Theta^{(2)}= \frac{\gamma}{\beta} \la F(x)^2\ra .
}

\subsection{Case of linear response regime}\lb{s:linear-theta}

The physical meaning of $\Theta^{(2)}$ is clear in the linear response regime.
We here refer to the word {\it linear response regime} to the situation that a system is attached to a single heat bath with $\beta$ and the probability distribution of the system is the canonical distribution with $\beta +\Di \beta$ ($\Di \beta \ll \beta$).
From a phenomenological viewpoint, the Fourier law $\Jq =\kappa \Di \beta$ holds with thermal conductance $\kappa$.
In addition, the entropy production is written as $\dsgm =\Jq \Di \beta$ up to $O(\Di \beta ^2)$.
By combining them, the inequality \eqref{Jsigma-2} suggests a relation $\kappa \leq \Theta^{(2)}$.
Interestingly, in the linear response regime the inequality turns out to be an equality:
\eqa{
\kappa =\Theta^{(2)}.
}{kappa-theta}
In other words, the inequality \eqref{Jsigma-2} is tight in the linear response regime.

We now derive \eqref{kappa-theta}.
In the rest of this section, we drop the dependence on time $t$, control parameter $\lmd (t)$ and label $\mu$, and neglect terms of $O(\Di \beta ^2)$.
Let $p^\beta _w:=e^{-\beta E_w}/Z^\beta$ be the canonical distribution with inverse temperature $\beta$, and we suppose that $p^\beta _w$ is invariant under the transition matrix $R$.
Using the relation ${\del Z^\beta}/{\del \beta}=-\la E\ra ^\beta Z^\beta$, we have
\eq{
p^{\beta+\Di \beta} _w-p^\beta _w=-p^\beta _w(E_w-\la E\ra ^\beta )\Di \beta,
}
where $\la \cdot \ra ^\beta$ represents the ensemble average with the canonical distribution with $\beta$.
The heat current $\Jq$ is then calculated as
\balign{
\Jq=&\sum_{w,w'}(E_w-E_{w'})R_{w'w}p^{\beta+\Di \beta}_w \nt \\
=&\sum_{w,w'}(E_w-E_{w'})R_{w'w}p^\beta _w(E_w-\la E\ra ^\beta )\Di \beta \nt \\
=&\frac{1}{2}\sum_{w,w'}(E_w-E_{w'})[R_{w'w}p^\beta _w(E_w-\la E\ra ^\beta )-R_{ww'}p^\beta _{w'}(E_{w'}-\la E\ra ^\beta )]\Di \beta \nt \\
=&\frac{1}{2}\sum_{w,w'}(E_w-E_{w'})^2R_{w'w}p^\beta _w\Di \beta \nt \\
=&\Theta^{(2)} \Di \beta ,
}
where we used $\sum_{w,w'}(E_w-E_{w'})R_{w'w}p^\beta _w=0$ in the second line and $R_{w'w}p^\beta _w=R_{ww'}p^\beta _{w'}$ in the fourth line.
The obtained inequality directly implies the desired relation \eqref{kappa-theta}.

\section{Application to heat engines}\lb{s:Markov-eff}

\subsection{Trade-off inequality between efficiency and power}

An important application of the inequalities \eqref{Jsigma-1} and \eqref{Jsigma-2} is to heat engines.
In this subsection, we shall derive a trade-off inequality between efficiency and power by applying the obtained inequality \eqref{Jsigma-1} or \eqref{Jsigma-2} to a cyclic process of a heat engine in $0\leq t\leq \tau$ (i.e., $\lmd(0)=\lmd (\tau)$) with two thermal baths with inverse temperatures $\beta_{\rm H}$ and $\beta_{\rm L}$ ($\beta_{\rm H}<\beta_{\rm L}$).
We assume that the initial and final probability distributions are the same: $p_{w,0}=p_{w,\tau}$, which embodies a cyclic process of heat engines\fn{
Here, one may feel that for a cyclic process of a macroscopic heat engine the initial and final states are considered to be the same only in the macroscopic sense, and microscopic probability distribution is not expected to be the same (i.e., $p_{w,0}=p_{w,\tau}$ is a non-realistic assumption for macroscopic engines).
However, fortunately, what we have utilized in our derivation is only the following weaker conditions that both the Shannon entropy and the energy expectation value are the same between the states at $t=0$ and $t=\tau$: 
\balign{
-\sum_w p_{w,0}\ln p_{w,0}=&-\sum_w p_{w,\tau} \ln p_{w,\tau} \\
\sum_w E_w^{\lmd (0)}p_{w,0}=&\sum_w E_w^{\lmd (\tau)}p_{w,\tau}.
}
Hence, if the Shannon entropy and energy do not change between the initial and the final states, our trade-off inequality \eqref{eff-power} is still valid even when other microscopic details are changed between the initial and the final states.
}.
We denote the hot and cold baths by H and L, respectively.
Then, 
\balign{
Q_{\rm H}:=&-\int_0^\tau dt\Jq _{\rm H}(t) \\
Q_{\rm L}:=&\int_0^\tau dt\Jq _{\rm L}(t)
}
represent the heat absorption from the hot bath and the heat emission to the cold bath, respectively.
The first law of thermodynamics implies that the work is expressed as $W=Q_{\rm H}-Q_{\rm L}$.

\

{\bf Theorem 2}:
In a cyclic process with two thermal baths, the power $W/\tau$ and efficiency $\eta :=W/Q_H$ satisfy
\eqa{
\frac{W}{\tau}\leq \bar{\Theta}\beta_{\rm L}\eta (\etac-\eta),
}{eff-power}
where $\bar{\Theta}:=\frac1\tau \int_0^\tau dt\Theta (t)$ represents the time-averaging of $\Theta (t)$ and $\etac :=1-\bH/\bL$ is the Carnot efficiency.

\

The inequality \eqref{eff-power} tells us that the power should vanish at $\eta =\etac$ and $\eta =0$.
The former is the desired result that the Carnot efficiency is attainable only with a quasistatic process.
The latter describes a trivial situation that we fail to extract any work (i.e., $W=0$).

The existence of a trade-off relation between efficiency and power has already been suggested in vast literature mainly on the basis of specific models and/or systems in the linear response regime~\cite{CA75, And84, SB12, BSS13, BS13, BBC13, BraS15, Yam16, BSS15, PB15, PCB16, PS17, Raz16}.
Notably, a trade-off inequality in the form $W/\tau \leq A \eta (\etac -\eta)$ ($A$: coefficient) has been obtained in Ref.~\cite{BSS15} for a periodically-driven underdamped Langevin system in the linear response regime.
For these backgrounds, a general trade-off inequality in this form has been expected to exist.

\bpf{
The increase of entropy\fn{
This entropy is defined in thermodynamic sense.
Because the entropy production in stochastic thermodynamics contains the entropy increase of baths as in the thermodynamic sense and the Shannon entropy of the systems is assumed to be invariant, this definition of $\Di S$ is consistent with the definition of entropy production in stochastic thermodynamics.
}
in the total system is given by
\eq{
\Di S=\bH\QH -\bL\QL ,
}
which is equal to $\int_0^\tau dt \dsgm (t)$ in a cyclic process.
Thus, the obtained inequality (\eqref{Jsigma-1} or \eqref{Jsigma-2}) provides a bound on $\Di S$ as
\balign{
(\QH +\QL)^2
&=\( -\int_0^\tau dt\Jq _{\rm H}(t)+\int_0^\tau dt\Jq _{\rm L}(t)\) ^2 \nt \\
&\leq \( \int_0^\tau dt \sum_\nu |\Jq _\nu(t)| \) ^2 \nt \\
&\leq \( \int_0^\tau dt \sqrt{\Theta (t)\dsgm (t)} \) ^2 \nt \\
&\leq \( \int_0^\tau dt {\Theta (t)} \) \(  \int_0^\tau dt \dsgm\)  \nt \\
&=\tau \bar{\Theta} \Di S,
}
where we used the Schwarz inequality in the fourth line.
Combining this inequality and a simple thermodynamic relation
\eq{
\eta (\etac-\eta)=\frac{W}{\QH}\( \frac{\QL}{\QH}-\frac{\bH}{\bL}\) =\frac{W}{\QH} \frac{\QL\bL-\QH\bH}{\bL\QH}=\frac{W\Di S}{\bL (\QH )^2},
} 
we arrive at the desired inequality
\eq{
\frac{W}{\tau}=\eta (\etac-\eta)\frac{\bL (\QH )^2}{\tau \Di S}\leq \eta (\etac-\eta)\frac{\bL (\QH )^2 \bar{\Theta}}{(\QH +\QL)^2}\leq \eta (\etac-\eta)\bL \bar{\Theta}.
}
}

\subsection{Remark on behavior of power near the Carnot efficiency}\lb{s:Carnot-power-detail}

The trade-off inequality between power and efficiency \eqref{eff-power} clearly exhibits the fact that finite power and the Carnot efficiency are incompatible as long as the coefficient $\Theta$ is finite, and \ref{s:fin-theta} confirms finiteness of $\Theta$.
However, finite power can exist near the Carnot efficiency.
In this subsection, we discuss possible unusual behavior of power near the Carnot efficiency, which should not be understood as the coexistence of finite power and the Carnot efficiency.

To illustrate the unusual behavior of power, we introduce a simple model which trivially realizes the apparent coexistence of finite power and the Carnot efficiency in a very loose sense.
The system consists of two states, 0 and 1, and the transition between 0 and 1 is induced by two heat baths, H and L.
The transition rates are set as
\balign{
R_{10}^{\rm H}&=ke^{-\bH (\Di E+F)/2}, \\
R_{01}^{\rm H}&=ke^{\bH (\Di E+F)/2}, \\
R_{10}^{\rm L}&=ke^{-\bL \Di E/2} , \\
R_{01}^{\rm L}&=ke^{\bL \Di E/2} ,
}
where $\Di E:=E_1-E_0$ is the energy difference, and $F$ is the external force coupling to the transition induced by the bath H.
We regard the transition against the external force $F$ as work.
If $\ep:=(\bL /\bH -1)\Di E-F>0$ is satisfied, the heat flows from the bath H to L in the stationary state and work is extracted steadily.
The stationary probability current from 0 to 1 via the transition with H is calculated as
\balign{
J_{0\to 1}^{\rm H}&=\frac{k(e^{\bH \ep/2}-e^{-\bH \ep/2})}{e^{-\bH (\Di E+F)/2}+e^{\bH (\Di E+F)/2}+e^{-\bL \Di E/2}+e^{\bL \Di E/2}} \nt \\
&=\frac{k\bH}{2(e^{-\bL \Di E/2}+e^{\bL \Di E/2})}\ep +O(\ep^2).
}

The limit $\ep\to 0$ leads to the efficiency $\eta =F/(\Di E+F)$ approaching to the Carnot efficiency $\etac =1-\bH/\bL$.
In the limit $\ep\to 0$ {\it with fixed $k$}, the stationary probability current converges to zero, which implies vanishing power.
On the other hand, if we take the limit $k\to \infty$ and $\ep\to 0$ {\it simultaneously} as satisfying $k\ep =\const$, then the stationary probability current remains at finite value, which implies finite power.
Moreover, if we take the limit $k\to \infty$ and $\ep\to 0$ with $k\ep ^2=\const$, then the power diverges and the efficiency approaches to the Carnot efficiency.

However, one should not consider that this model is an example of the coexistence of finite power and the Carnot efficiency.
This is because the coefficient $k$ reflects the inherent time-scale of the system and changing $k$ means changing the time-scale of the system.
What we say in ``coexistence of finite power and the Carnot efficiency" is that both finite power and the Carnot efficiency realize {\it with keeping the time-scale of the system}.

On the basis of these observations, we may regard the coefficient $\Theta$ as a kind of a time-scale parameter of energy exchange of the system.
Our result claims that the only possible way to increase the power with keeping high efficiency is the trivial improvement as explained above, and other nontrivial improvements do not exist.
The model of Polettini and Esposito~\cite{PE17} can be understood as showing this point from the opposite perspective.

\section{Entropy production inequality for general quantities}\lb{s:Markov-genform}

\subsection{General bound}

In the derivations of Eqs.\eqref{Jsigma-1} and \eqref{Jsigma-2}, we have not used the fact that the current $\Jq$ is the heat current.
In fact, the entropy production bounds the time derivative of any quantity $G$.

\

{\bf Theorem 3.1}:
If the canonical distribution is invariant, the entropy production rate bounds the time derivative of any quantity $G_w^{\lmd (t)}$:
\eqa{
\sum_{\nu =1}^k |J^G_\nu (t)|\leq \sqrt{\Theta^{G1} (t)\dsgm (t)}
}{GJsigma-1}
with
\balign{
J^G_\nu (t):=&-\sum_{w,w'} G_w^{\lmd (t)} R_{ww'}^{\mu, \lmd (t)}p_{w',t} \\
\Theta^{G1} (t):=&\frac{1}{c_0}\sum_\mu \sum_{w\neq w'}(\Di G_w^{\mu, \lmd (t)})^2 (R_{ww'}^{\mu, \lmd (t)}p_{w',t}+R_{w'w}^{\mu, \lmd (t)}p_{w,t}).
}
Here, $\Di G_w^{\mu, \lmd (t)}:=G_w^{\lmd (t)}-\la G^{\lmd (t)}\ra_{t, w^{-i}}$ represents the fluctuation of $G$ under the condition that all particles except the $i$-th one are fixed.
If there is only a single bath, then $J^G_\nu (t)=d\la G\ra/dt$.

\

{\bf Theorem 3.2}:
If the system satisfies the local detailed-balance condition~\eqref{db}, we have
\eqa{
\sum_{\nu =1}^k |J^G _\nu (t)|\leq \sqrt{\Theta^{G2} (t)\dsgm (t)}
}{GJsigma-2}
with
\eq{
\Theta^{G2} (t):=\frac{1}{2}\sum_{w\neq w'}(G_w^{\lmd (t)}-G_{w'}^{\lmd (t)})^2 R_{ww'}^{\lmd (t)}p_{w',t}.
}

\

These relations are proved by replacing $E_w$ (in $\Jq$ and $\Theta$) by $G_w$ in the derivations of Eqs.\eqref{Jsigma-1} and \eqref{Jsigma-2}.

Notably, $G$ is not assumed to be a conserved quantity.
We, however, remark that for the case with Hamiltonian dynamics and non-conserved $G$, our trick to remove the effect of Hamiltonian dynamics seen in \sref{Lan-theta} no longer works.
In general, $\Theta^{G1}$ for Hamiltonian dynamics diverges in the continuum limit.

\subsection{Case of thermoelectric transport}\lb{s:thermoele}

Applying the obtained inequality to thermoelectric transport, we have a similar trade-off relation to \eref{eff-power} between power and efficiency.
Consider two heat-particle baths with inverse temperatures and chemical potentials $\beta_1, \mu_1$ and $\beta_2, \mu _2$ satisfying $\beta_1<\beta_2$ and $\mu_1<\mu_2$~\cite{BCSW}.
The heat and particle currents from the bath 1 to the bath 2 are denoted by $\Jq$ and $\Jn$, both of which we assume positive.
The efficiency of thermoelectricity is defined as
\eq{
\eta :=\frac{\Di \mu \Jn}{\Jq -\mu_1\Jn}
}
with $\Di \mu :=\mu_2-\mu_1>0$, where we also assumed $\Jq -\mu_1\Jn>0$ (see also \ref{s:eff-te}).
We note that we defined the power (work extraction per unit time) by $\Di \mu \Jn$, not by $(\beta_2 \mu_2-\beta_1\mu_1)\Jn$.

The inequalities \eqref{GJsigma-1} and \eqref{GJsigma-2} in the previous subsection suggest the following trade-off relations for heat and particle currents:
\balign{
2\Jq \leq& \sqrt{\Theta^{\rm q}\dsgm}, \\
2\Jn \leq& \sqrt{\Theta^{\rm n}\dsgm}.
}
Here, $\Theta^{\rm q}$ is $\Theta^{(1)}$ or $\Theta^{(2)}$ defined in \eref{theta1-def} or \eref{theta2-def}, and $\Theta^{\rm n}$ is defined as
\balign{
\Theta^{\rm 1,n}:=&\frac{1}{c_0}\sum_\mu \sum_{w\neq w'}(\Di N_w^{\mu})^2 (R_{ww'}^{\mu}p_{w'}+R_{w'w}^{\mu, \lmd (t)}p_{w}) \\
\intertext{or}
\Theta^{\rm 2,n}:=&\frac{1}{2}\sum_{w\neq w'}(N_w-N_{w'})^2 R_{ww'}p_{w'},
}
where $N_w$ represents the number of particles in the state $w$.

\

{\bf Theorem 4}:
In thermoelectric transport, the power $\Di \mu \Jn$ and efficiency $\eta$ satisfy
\eq{
\Di \mu \Jn \leq \frac{\Theta^{\rm q}+\mu_1^2 \Theta^{\rm n}}{2} \beta_{\rm 2}\eta (\etac-\eta).
}

\

\bpf{
First, $\eta (\etac -\eta )$ is calculated as
\balign{
\eta (\etac -\eta )
=&\frac{\Di \mu \Jn}{\Jq -\mu_1\Jn} \( 1-\frac{\beta_1}{\beta_2}- \frac{\Di \mu \Jn}{\Jq -\mu_1\Jn}\) \nt \\
=&\frac{\Di \mu \Jn}{\Jq -\mu_1\Jn} \frac{(\beta_2-\beta_1)(\Jq-\mu_1\Jn )- \beta_2 (\mu_1-\mu_2)\Jn}{\beta_2 (\Jq -\mu_1\Jn)} \nt \\
=&\frac{\Di \mu \Jn \cdot \dsgm}{\beta_2 (\Jq -\mu_1\Jn)^2}.
}
Then, the inequalities between the current and the entropy production rate suggest
\balign{
\frac{(\Theta^{\rm q}+\mu_1^2 \Theta^{\rm n}) \dsgm}{2}
&\geq 2((\Jq)^2+\mu_1^2 (\Jn)^2)\geq  (\Jq)^2+\mu_1^2 (\Jn)^2 + 2|\Jq\mu_1\Jn|\geq (\Jq -\mu_1 \Jn )^2,
}
where we used a relation $(a+b)/2\geq \sqrt{ab}$ in the second line.
Combining these two relations, we arrive at the desired inequality.
}

\section{Quantum case}\lb{s:quantum}
In this section, we briefly address a quantum case, where the dynamics is described by the Lindblad equation \cite{bptext}. Let $\rho (t)$ and $H_S(t)$ be the density matrix of the system and the system's Hamiltonian at time $t$. 
In previous studies, although stationary Lindblad systems~\cite{SS16} and quantum processes described with the microscopic viewpoint~\cite{ST17, Llo18} have been investigated, the general Lindblad dynamics has not yet been addressed.
We demonstrate that the quantum version of the relations \eqref{Jsigma-1-simple} and (\ref{Jsigma-2-simple}) can be derived.
The idea of this derivation is inspired by Ref.~\cite{funo}.

We first consider the case with the local detailed-balance condition.
We start with the general expression of the Lindblad equation described as follows \cite{bptext}
\begin{align}
{\partial \rho (t) \over \partial t} &= -{i \over \hbar} \left[ H_S (t) , \rho (t) \right] + {\cal D}\left[ \rho (t) \right] \, , \\
 {\cal D}\left[ \rho (t) \right]  &= \sum_{a,\xi }\gamma_a (\xi ) 
 \left\{ 
 L_{a,\xi} (t) \rho (t)  L_{a,\xi}^{\dagger} (t)  -{1\over 2}  L_{a,\xi}^{\dagger} (t) L_{a,\xi} (t)  \rho (t)  -{1\over 2} \rho (t)  L_{a,\xi}^{\dagger} (t) L_{a,\xi} (t) 
  \right\}  ,
\end{align}
where $a$ is the index of the operator $L_a$, and the operator $L_{a,\xi}$ is defined based on the operator $L_a$ though the eigenstates of the Hamiltonian $H_S(t)$:
\begin{align}
L_{a,\xi} &:= \sum_{\epsilon - \epsilon'=\xi}{\rm \Pi}_{\epsilon'} L_a {\rm \Pi}_{\epsilon} \, . 
\end{align}
Here, ${\rm \Pi}_{\epsilon}:=\ket{\ep}\bra{\ep}$ is the projection operator onto the eigenstates with the eigenvalue $\epsilon$. 
The local detailed-balance condition is expressed by 
\eq{
\frac{\gamma_a (-\xi)}{\gamma_a (\xi)} = e^{-\beta_a \xi}.  
}

Let $|n (t) \rangle$ be an eigenstate of the density matrix $\rho (t)$ satisfying $\rho (t) |n (t) \rangle =p_{n} (t) | n(t) \rangle $. Then, one can derive the following expression for the heat current:
\begin{align}
J^q (t) &= -{\rm Tr}\left( H_S (t) {\cal D}\left[ \rho (t) \right]\right) =\sum_{a,\xi}\sum_{n,m} \xi \gamma_a(\xi )|\langle n (t) | L_{a,\xi} (t) | m(t) \rangle |^2 p_m (t) \, . 
\end{align}
Similarly, the entropy production rate of the system is given by
\begin{align}
{d \over dt} H (\rho (t)) &= -{\rm Tr}\left(  {\cal D}\left[ \rho (t) \right] \log \rho (t) \right) \nonumber \\
&= \sum_{a,\xi}\sum_{n,m}  \gamma_a(\xi ) |\langle n (t) | L_{a,\xi} (t) | m(t) \rangle |^2
p_m (t) \log (p_m(t)/p_n(t) ) \, . 
\end{align}
Now we define the following quantity
\begin{align}
R^{a,\xi}_{n,m}  (t) &:=\gamma_a(\xi ) |\langle n (t) | L_{a,\xi} (t) | m(t) \rangle |^2 \, .
\end{align}
Remark that this matrix satisfies the normalization condition in the following sense:
\eq{
\sum_{a,\xi, n}R^{a,\xi}_{n,m}  (t)=0.
}
Then the heat current and the total entropy production rate are respectively written as follows \cite{funo}
\begin{align}
\Jq (t)&= \sum_{a,\xi}\sum_{n,m}\xi R^{a,\xi}_{n,m} (t) p_m (t) = {1\over 2}\sum_{a,\xi}\sum_{n,m} \xi ( R^{a,\xi}_{n,m} (t) p_m (t) -R^{a,-\xi}_{m,n} (t)  p_n (t) ) \, , \\
\dot{\sigma}(t) &=\sum_{a,\xi}\sum_{n,m} R^{a,\xi}_{n,m} (t)  p_m (t) \log \left( { R^{a,\xi}_{n,m} (t) p_m (t) \over R^{a,-\xi}_{m,n} (t) p_n (t)   }\right) \, .
\end{align}
This structure is analogous to the classical case. Hence following the procedure explained in \sref{Markov-db}, we can immediately obtain the following relation
\begin{align}
|\Jq (t) | &\le \sqrt{\Theta_{\rm q}^{(2)} \dot{\sigma} (t)} \, , \\
\Theta_{\rm q}^{(2)} &:= {1\over 2} \sum_{a,\xi}\sum_{n \neq m} \xi^2 R^{a,\xi}_{n,m} (t) p_m (t) \, . 
\end{align}

\

We nest consider more general Lindblad dynamics.
To consider the canonical distribution with no ambiguity, we here assume that the system Hamiltonian $H_S(t)$ has no degeneracy.
The requirement of invariance of the canonical distribution is expressed as
\eqa{
\sum_{a,\ep}\gamma_a(\ep-\ep')\abs{\bra{\ep'}L_a(t)\ket{\ep}}^2e^{-\beta_a \ep}=0
}{q-can-inv}
for any $\ep'$.
We now introduce the dual matrix of $R^{a,\xi}_{m,n}$ defined as
\eqa{
\tlr ^{a,-\xi}_{m,n}:=R^{a,\xi}_{n,m}e^{-\beta_a \xi} = \gamma_a(\xi ) |\langle n (t) | L_{a,\xi} (t) | m(t) \rangle |^2 e^{-\beta_a \xi}.
}{q-dual}
The dual matrix can be regarded as a quantum Markov process with
\balign{
\tilde{\gamma}_a(\xi)=&\gamma_a(-\xi)e^{\beta_a \xi}, \\
\tilde{L}_{a,\xi}=&L^\dagger_{a,-\xi}.
}

It might be useful to define the transition matrix
\eq{
R^a_{(\ep', m)(\ep,n)}:= \gamma_a(\ep-\ep' ) \abs{\braket{m(t)|\ep'} \bra{\ep'}L_a(t)\ket{\ep} \braket{\ep|n(t)}}^2
}
and its dual matrix
\eq{
\tlr^a_{(\ep,n)(\ep',m)}:=\gamma_a(\ep'-\ep ) \abs{ \braket{n(t)|\ep} \bra{\ep}L_a(t)\ket{\ep'} \braket{\ep'|m(t)}}^2 e^{-\beta_a (\ep' -\ep)}.
}
Notably, even though we have {\it not} assumed the no resonance condition (i.e., $\ep-\ep'=\ep''-\ep'''$ implies $\ep=\ep''$ and $\ep'=\ep'''$), we obtain the following relation
\eqa{
\sum_{n}R^{a,\xi}_{m,n}=\sum_{\substack{n, \ep, \ep' \\ \ep-\ep'=\xi}}R^a_{(\ep', m)(\ep,n)}.
}{noresonancelike}
This relation is shown as follows:
\balign{
\sum_{n}R^{a,\xi}_{m,n}
=&\sum_{n}  \gamma_a(\xi ) |\langle m (t) | L_{a,\xi} (t) | n (t) \rangle |^2  \nt \\
=&\sum_{n}  \gamma_a(\xi) \abs{ \sum_{\substack{\ep, \ep' \\ \ep-\ep'=\xi}} \braket{m(t)|\ep'} \bra{\ep'}L_a(t)\ket{\ep} \braket{\ep|n(t)}}^2  \nt \\
=&\sum_{n}  \gamma_a(\xi) \sum_{\substack{\ep, \ep', \ep'', \ep''' \\ \ep-\ep'=\xi, \ep''-\ep'''=\xi}} \braket{m(t)|\ep'} \bra{\ep'}L_a(t)\ket{\ep} \braket{\ep|n(t)} \braket{n(t)|\ep''} \bra{\ep''}L_a^\dagger (t)\ket{\ep'''} \braket{\ep'''|m(t)}  \nt \\
=&\gamma_a(\xi) \sum_{\substack{\ep, \ep', \ep'', \ep''' \\ \ep-\ep'=\xi, \ep''-\ep'''=\xi}} \braket{m(t)|\ep'} \bra{\ep'}L_a(t)\ket{\ep} \braket{\ep|\ep''} \bra{\ep''}L_a^\dagger (t)\ket{\ep'''} \braket{\ep'''|m(t)}  \nt \\
=&\gamma_a(\xi) \sum_{\substack{\ep, \ep' \\ \ep-\ep'=\xi}} \braket{m(t)|\ep'} \bra{\ep'}L_a(t)\ket{\ep} \bra{\ep}L_a^\dagger (t)\ket{\ep'} \braket{\ep'|m(t)}   \nt \\
=&\gamma_a(\xi) \sum_{\substack{\ep, \ep' \\ \ep-\ep'=\xi}} \abs{\braket{m(t)|\ep'} \bra{\ep'}L_a(t)\ket{\ep}}^2 \sum_n \abs{\braket{\ep|n(t)}}^2   \nt \\
=&\sum_{\substack{n, \ep, \ep' \\ \ep-\ep'=\xi}}R^a_{(\ep', m)(\ep,n)}.
}
In the fifth line we used the condition of no degeneracy (i.e., $\ep-\ep'=\ep-\ep'''=\xi$ only if $\ep'=\ep'''$).
In a similar manner to above, we can show a similar relation for the dual transition matrix
\eq{
\sum_{m}\tlr^{a,-\xi}_{n,m}=\sum_{\substack{n, \ep, \ep' \\ \ep'-\ep=-\xi}}\tlr^a_{(\ep,n)(\ep',m)}.
}
Using \eref{noresonancelike}, the normalization condition for the dual transition matrix is easy to obtain as follows:
\balign{
\sum_{a, m,\xi} \tlr ^{a,-\xi}_{m,n}&=\sum_{a, m,\xi} R^{a,\xi}_{n,m}e^{-\beta_a \xi} \nt \\
&=\sum_{a, m, \ep, \ep'}R^a_{(\ep, n)(\ep',m)}e^{-\beta_a (\ep'-\ep)}  \nt \\
&=\sum_{\ep} \abs{\braket{n(t)|\ep}}^2 e^{\beta_a\ep} \sum_{a,\ep'}\gamma_a(\ep-\ep')\abs{\bra{\ep}L_a(t)\ket{\ep'}}^2e^{-\beta_a \ep'} \nt \\
&=0,
}
where in the third line we used a relation $\sum_m \abs{\braket{\ep'|m(t)}}^2=1$, and in the last line we used the invariance of the canonical distribution \eqref{q-can-inv}.

Using these, the heat current is written as
\balign{
\Jq (t)=&\sum_{a} \sum_{m} \sum_{\ep, \ep' } (\ep-\ep') R^a_{(\ep',m)(\ep,n)}p_n(t)  
=-\sum_a \sum_{m,n}\sum_{\ep.\ep'} \ep' R^a_{(\ep',m)(\ep, n)} p_n(t).
}
In the second equality, we used the normalization condition $\sum_{a,\ep',m}R^a_{(\ep',m)(\ep,n)} =0$ for any $n$ and $\ep$.
Thus, the heat current and the entropy production rate is written as 
\balign{
\Jq (t)&=-\sum_{a,\ep, \ep'}\sum_{n,m}\Di \ep' (t) R^{a}_{(\ep', m)(\ep,n)}(t) p_n (t) \nt \\
&=-\sum_{a,\ep,\ep'}\sum_{n,m} \Di \ep' (t) \[ R^{a}_{(\ep', m)(\ep,n)}(t) p_n (t) -\tlr^{a}_{(\ep,n)(\ep',m)} (t)  p_m (t) \] , \\
\dsgm (t)&=\sum_{a,\ep,\ep'}\sum_{n,m} R^{a}_{(\ep',m)(\ep,n)} (t)  p_n (t) \log \left( {R^{a}_{(\ep',m)(\ep,n)} (t)p_n (t) \over \tlr^{a}_{(\ep,n)(\ep',m)} (t) p_m (t)   }\right).
}
Here, we defined $\Di \ep'(t)$ as
\eq{
\Di \ep'(t):=\ep' -\Tr [H_S(t)\rho(t)].
}
Then, following the procedure in \sref{Markov-gen}, we obtain the trade-off inequality for general quantum Markov processes:
\balign{
|\Jq (t) | &\le \sqrt{\Theta_{\rm q}^{(1)} \dot{\sigma} (t)} \, , \\
\Theta_{\rm q}^{(1)} &:= c_0 \sum_{a,\ep,\ep'}\sum_{n\neq m} (\Di \ep')^2 \[ R^{a}_{(\ep', m)(\ep,n)}(t) p_n (t) + R^{a}_{(\ep,n)(\ep',m)} (t)  p_m (t) \] \, . 
}

\section{Discussion}

We have derived trade-off inequalities between entropy production and heat current.
Our result is applicable to any classical and quantum Markovian systems including systems with broken time-reversal symmetry in transient processes with a time-dependent Hamiltonian. 
The wide applicability of our inequalities, in particular for time-dependent systems, comes from the fact that our inequalities treat instantaneous quantities only.
The obtained inequality \eqref{Jsigma-2} is tight in the linear response regime, in which the coefficient $\Theta$ becomes thermal conductivity.
As the corollary of the main inequalities, we obtained a no-go theorem that finite power and the Carnot efficiency are incompatible.

The crucial idea in our derivations is the decomposition of the entropy production rate.
As explained before, our proof is inspired by the idea of partial entropy production, which is decomposition of entropy production.
The decomposition also plays an important role to improve $\Theta$ such that the inequalities are meaningful.
In \eref{R-decomp}, we have introduced the decomposition of the time-evolution operator $R$ into the contribution of Hamiltonian dynamics and those of stochastic dynamic of each particle with each bath.
As shown in \ref{s:thermo-theta}, the decomposition into each particle keeps $\Theta^{(1)}$ finite in the thermodynamic limit.
In addition, owing to this decomposition the stronger inequality \eqref{Jsigma-2} is applicable to underdamped Langevin systems by removing the effect of Hamiltonian dynamics.
In fact, this procedure removes all effects from a field with broken time-reversal symmetry (e.g., Lorentz force), potential energy (including both interaction energy and one-body potential energy) dependent on their positions, and inertia acting on their positions.
The remaining time-evolution operator acts only on the momentum of a single particle, which satisfies the local detailed-balance condition.

We here remark that our result strongly relies on the Markov property, and thus it seems to be not easy to extend our results and techniques to non-Markovian systems.
Some attempts to derive trade-off relations on speed and efficiency are seen in Refs.~\cite{ST17, Llo18}.

It is worth comparing our result to the thermodynamic uncertainty relation~\cite{BarS15, Ging16, GRH17, PRS17, HG17, DS17}, which connects fluctuation of time-integrated heat current and entropy production in a very similar form to our result.
The thermodynamic uncertainty relation was first found for the case of the long-time limit~\cite{BarS15, Ging16, GRH17}, and then extended to the case of a finite time interval~\cite{PRS17, HG17, DS17}.
We first emphasize that these two relations concern different quantities:
Our result considers instantaneous quantities, while the thermodynamic uncertainty relation considers time-integrated quantities.
However, considering some limiting cases, we can compare these two results directly.
The short time interval limit of the finite-time thermodynamic uncertainty relation~\cite{PRS17} reproduces our inequality \eqref{Jsigma-2} with the local detailed-balance condition.
In addition, our inequality can be extended to time-integrated quantities for stationary systems with the local detailed-balance condition as shown in \ref{s:relax-tur}.
The obtained inequality is weaker than the thermodynamic uncertainty relation.
On the other hand, the thermodynamic uncertainty relation applies only to specific setups; stationary systems with the local detailed-balance condition described by continuous-time Markov jump processes.
The thermodynamic uncertainty relation no longer holds in systems with one of these conditions violated including systems with momentum or a magnetic field~\cite{BHS17, MBG18}, transient processes and relaxation processes (see \ref{s:relax-tur}), and discrete-time Markov chain processes~\cite{Shi17-note, PB17}.
By contrast, our approach presented in this paper is applicable to a non-stationary system with time-dependent transition rate, a system with parity-odd fields or variables, and a Markov chain, which is the advantage of our result.

Closing this paper, we put a remark on the coefficient $c_0=8/9$ in $\Theta^{(1)}$.
As shown in \ref{s:best-coeff}, $c_0=8/9$ is not a tightest coefficient, and the best coefficient of $\Theta^{(1)}$ is numerically calculated as $c^*= 0.89612\cdots$, which has been appeared in some literatures~\cite{SST16, BHS17, SFS18}.
One may feel that this coefficient embodies only the limitation of our approach and this quantity is physically meaningless.
However, maybe surprisingly, a numerical simulation reveals that a variants of the thermodynamic uncertainty relation under a magnetic field indeed has the same coefficient $c^*= 0.89612\cdots $ as an achievable bound~\cite{BHS17}.
Although the form of the inequality considered in Ref.~\cite{BHS17} is slightly different from our inequality \eqref{Jsigma-1}, this fact strongly suggests that the coefficient $c^*= 0.89612\cdots $ indeed reflects the physics of our world.

\begin{acknowledgements}
We are grateful to Hal Tasaki for fruitful discussion.
He was a co-author in the joint work~\cite{SST16}, and contributed to deriving several relations.
NS was supported by Grant-in-Aid for JSPS Fellows JP17J00393. 
KS was supported by JSPS Grants-in-Aid for Scientific Research (No. JP25103003, JP16H02211 and JP17K05587).

\end{acknowledgements}

\appendix

\renewcommand{\thesection}{Appendix.\Alph{section}}

\section{Analysis with linear irreversible thermodynamics}\lb{s:linear}
 \appnum{\Alph{section}}

In this appendix, we clarify the fact that if time-reversal symmetry is broken the linear irreversible thermodynamics does not prohibit the existence of a heat engine with the Carnot efficiency at finite power~\cite{BSC11}.
We consider a stationary system with two kinds of flux $J_1$ and $J_2$ with corresponding thermodynamic forces $X_1$ and $X_2$.
We set $J_2$ to heat flux and $X_2:=1/T_L-1/T_H$, $J_1$ to another flux which flows against the thermodynamic force $X_1$ (i.e., $X_1J_1\leq 0$ and $X_2J_2\geq 0$).
The power and efficiency are given by $\dot{W}:=-X_1J_1T$ and $\eta:=-X_1J_1T/J_2$.

We consider a system with a magnetic field $B$.
In this system, the linear expansion of the flux $J$ is written as
\balign{
J_1=&L_{11}(B)X_1+L_{12}(B)X_2, \\
J_2=&L_{21}(B)X_1+L_{22}(B)X_2,
}
where $L$ is the Onsager matrix.
The Onsager reciprocity relation tells $L_{12}(B)=L_{21}(-B)$, and in general $L_{12}(B)\neq L_{21}(B)$.
In the remainder of this appendix, since we consider only systems with a magnetic field $B$, we omit the parameter $B$.
The entropy production rate $\dot{S}:=J_1X_1+J_2X_2$ is calculated as
\balign{
\dot{S}=&L_{11}X_1^2+(L_{12}+L_{21})X_1X_2+L_{22}X_2^2 \nt \\
=&L_{11}\( X_1+\frac{L_{12}+L_{21}}{2L_{11}}X_2\) ^2 +\( L_{22}-\frac{(L_{12}+L_{21})^2}{4L_{11}}\) X_2^2. \lb{linear-dotS}
}
Because the second law of thermodynamics claims $\dot{S}\geq 0$ for any $X_1$ and $X_2$, by setting $X_1=-(L_{12}+L_{21})X_2/2L_{11}$, we find that the coefficient of the second term of \eref{linear-dotS} is nonnegative: 
\eq{
L_{22}-\frac{(L_{12}+L_{21})^2}{4L_{11}}\geq 0.
}
This condition suggests that the entropy production rate is bounded by a quadratic term:
\eqa{
\dot{S}\geq L_{11}\( X_1+\frac{L_{12}+L_{21}}{2L_{11}}X_2\) ^2 =\frac{1}{L_{11}}\( J_1+\frac{L_{21}-L_{12}}{2}X_2\) ^2.
}{dotS-ineq}

We now investigate the condition for the Carnot efficiency $\dot{S}=0$.
We first consider the case with time-reversal symmetry (i.e., $L_{12}=L_{21}$).
In this case, \eref{dotS-ineq} reduces to
\eq{
\abs{J_1}\leq \sqrt{\dot{S}L_{11}},
}
which looks very similar to \eref{Jsigma-1}, and clearly shows that the Carnot efficiency $\dot{S}=0$ is achievable only when power is zero: $\abs{J_1}=0$.

We next consider the case without time-reversal symmetry.
Equation \eqref{linear-dotS} suggests that $\dot{S}=0$ holds if and only if the following two conditions
\balign{
L_{22}-\frac{(L_{12}+L_{21})^2}{4L_{11}}&=0 \lb{dotS-cond1} \\
J_1+\frac{L_{21}-L_{12}}{2}X_2=L_{11}X_1+\frac{L_{21}+L_{12}}{2}X_2&=0 \lb{dotS-cond2}
}
are satisfied simultaneously.
Then, if $L_{12}\neq L_{21}$, for any $L$ satisfying $L_{22}-{(L_{12}+L_{21})^2}/{4L_{11}}=0$ and any nonzero $X_2$, there exists nonzero $X_1=-{(L_{21}+L_{12})X_2}/{2L_{11}}$ satisfying \eref{dotS-cond2}.
We note that $X_2\neq 0$ and $L_{21}-L_{12}\neq 0$ directly imply finite power: $J_1\neq 0$.
Since the second condition \eqref{dotS-cond2} can be always satisfied by setting nonzero $X_1$ and $X_2$ properly as long as $L_{12}\neq L_{21}$, the remaining question is whether the first condition \eqref{dotS-cond1} is realizable under $L_{12}\neq L_{21}$.
However, within the framework of the linear irreversible thermodynamics, there is no {\it a priori} reason to exclude the possibility of $L_{22}-{(L_{12}+L_{21})^2}/{4L_{11}}=0$ with $L_{12}\neq L_{21}$.

\

This clearly shows that finite power and the Carnot efficiency is compatible under a magnetic field.
We, however, should note that the above analysis only shows that the linear irreversible thermodynamics does not formally exclude the possibility of a heat engine with the Carnot efficiency at finite power, and does not show that there indeed exists such a heat engine.
In fact, as seen in the main part of this paper, by taking into account microscopic details of the system, we find that the Carnot efficiency and finite power are incompatible.

\section{Discretization and continuum limit of Kramers equation and Hamilton's equation}\lb{s:discre}
 \appnum{\Alph{section}}

In this Appendix, we provide the detailed procedure of the discretization and continuum limit for continuous systems, which is briefly discussed in \sref{discre-main}.
Same as \sref{discre-main}, we consider a Markov process of a single particle in one-dimensional continuous space described by the following Kramers equation:
\eqa{
\frac{d}{dt}P( x ,  p)= \[ -\frac{ p}{m} \cdot \frac{\del}{\del  x} +\frac{\del}{\del  p} \cdot \( \frac{\gamma  p}{m} - F(x,p) \) +\frac{\gamma}{\beta }\frac{\del^2}{\del { p}^2}  \] P( x ,  p),
}{FP}
where $ x$ and $ p$ are the position and momentum of the particle.
We remark that stochastic Markov processes obtained through the system size expansion always take this form of equation~\cite{Kambook}.

The right-hand side of \eref{FP} is decomposed into the Hamiltonian part
\eqa{
\[ -\frac{ p}{m} \cdot \frac{\del}{\del  x} - \frac{\del}{\del  p} \cdot F(x,p) \] P(x,p)
}{Ham-part}
and the dissipative part
\eqa{
\[ \frac{\del}{\del  p} \cdot \frac{\gamma  p}{m} +\frac{\gamma}{\beta }\frac{\del^2}{\del { p}^2}\] P(x,p).
}{dissip-part}
The former is equivalent to Hamilton's equation:
\balign{
\frac{d}{dt}p=&F(x,p) \\
\frac{d}{dt}x=&\frac{p}{m}.
}
The latter is equivalent to the following Langevin equiation:
\eqa{
\frac{d}{dt}p=-\frac{\gamma}{m}p+\sqrt{\frac{2\gamma}{\beta}}\xi (t).
}{p-Langevin}
Here, $\xi (t)$ represents the white Gaussian noise.
The first term represents the viscous resistance, and the second term represents stochastic thermal noise.
The equivalence of the Langevin equation and the Fokker-Planck equation is shown in many textbooks~\cite{Kambook}.

\

We first consider the discretized transition matrix corresponding to the dissipative part.
The transition matrix from a state with momentum $p$ to $p\pm \ep$ is given by \eref{disc-rate-main}, which reappears below:
\eqa{
{R}_{p\pm \ep ,p}=\frac{\gamma}{\beta \ep^2}e^{-\frac{\beta}{4m}((p\pm \ep )^2-p^2)}=\frac{\gamma}{\beta \ep^2} e^{O(\ep)}.
}{disc-rate}
We shall show that this transition rate indeed reproduces the dissipative part \eqref{dissip-part}.
Expanding the transition matrix in $\ep$ as
\balign{
R_{p\pm \ep,p}=&A\( 1\mp \frac{\beta}{2m}\ep p+\frac{\beta}{4m}\ep^2+\frac{\beta^2}{8m}\ep^2p^2 +O(\ep^3) \)  \\
R_{p, p\pm \ep}=&A\( 1\pm \frac{\beta}{2m}\ep p-\frac{\beta}{4m}\ep^2+\frac{\beta^2}{8m}\ep^2p^2 +O(\ep^3) \) 
}
with $A:=\gamma/\beta \ep^2$, the master equation with \eref{disc-rate} becomes
\balign{
\frac{d}{dt}P(p)
=& -(R_{p+\ep,p}+R_{p-\ep,p})P(p)+R_{p,p+\ep}P(p+\ep)+R_{p,p-\ep}P(p-\ep) \nt \\
=&A\( 1+\frac{\beta^2}{8m}\ep^2p^2\) (P(p+\ep)+P(p-\ep)-2P(p)) \nt \\
&+A\frac{\beta}{2m}\ep p (P(p+\ep)-P(p-\ep)) \nt \\
&+A\frac{\beta}{4m}\ep^2(P(p+\ep)+P(p-\ep)+2P(p)) +O(\ep).
}
Taking $\ep \to 0$ limit, we recover the Kramers equation:
\eq{
\frac{d}{dt}P(p)=\frac{\del}{\del p} \( \frac{\gamma p}{m}P(p)\) +\frac{\gamma}{\beta}\frac{\del^2}{\del p^2} P(p).
}
Hence, the discretization with the transition rate \eqref{disc-rate} indeed reproduces the time-evolution of \eqref{dissip-part}.

\

We next consider the discretized transition matrix corresponding to the Hamiltonian part.
This discretization draws the $p-x$ phase space as the $\ep \times \ep '$ lattice.
A single state is determined by a pair of position and momentum, $(x,p)$.
Supposing $p>0$ and $F(x,p)>0$, we set the transition matrix of $(x,p)$ as Eqs.~\eqref{disc-det1-main} and \eqref{disc-det2-main}, which reappear below:
\balign{
R_{(x,p+\ep),(x,p)}&:=\frac{1}{\ep}F(x,p), \lb{disc-det1} \\
R_{(x+\ep',p),(x,p)}&:=\frac{1}{\ep'}\frac{p}{m}.\lb{disc-det2}
}
We remark that the inverse transitions of the above transitions do not occur (i.e., $R_{ (x,p),(x,p+\ep)}=0$ and $R_{(x,p),(x+\ep',p)}=0$).
The master equation reads
\balign{
\frac{d}{dt}P(x,p)
=&-P(x,p)(R_{(x,p+\ep),(x,p)}+R_{(x+\ep',p),(x,p)}) \nt \\
&+P(x,p-\ep)R_{(x,p),(x,p-\ep)}+P(x-\ep',p)R_{(x,p),(x-\ep',p)} \nt \\
=&\frac{1}{\ep'}\frac{p}{m}(P(x,p-\ep')-P(x,p))+\frac{1}{\ep}(F(x,p-\ep)P(x,p-\ep)-F(x,p)P(x,p)),
}
whose continuum limit $\ep, \ep' \to 0$ reproduces the Liouville operator
\eq{
\frac{d}{dt}P(x,p)=-\frac{p}{m}\frac{\del}{\del x}P(x,p)-\frac{\del}{\del p}F(x,p)P(x,p).
}
Hence, the discretization with the transition rates \eqref{disc-det1} and \eqref{disc-det2} indeed reproduce the time-evolution of \eqref{Ham-part}.

\section{Proof of Lemma 1}\lb{s:lemma1}
\appnum{\Alph{section}}

We recast the Lemma 1:
\eq{
D(p||q)\geq c_0 \sum_i \frac{(p_i-q_i)^2}{p_i+q_i}
}
with $c_0=8/9$.
In the following, we shall show the proof of this inequality.

\

\bpf{
We first show an inequality
\eqa{
a\ln \frac{a}{b}+b-a\geq c_0 \frac{(a-b)^2}{a+b}
}{89ineq}
for any $a,b >0$.
This inequality is equivalent to
\eq{
\frac{1}{a}\( a\ln \frac{a}{b}+b-a- c_0 \frac{(a-b)^2}{a+b}\) =-\ln u +u-1-\frac{c_0 (1-u)^2}{1+u}=:h(u)\geq 0
}
with $u:=b/a$.
Since $h(1)=0$, it is enough to show that the derivative of $h(u)$
\eq{
h'(u)=\frac{u-1}{u(1+u)^2}\{ (1-c_0)u^2+(2-3c_0)u+1\}
}
satisfies $h'(u)\geq 0$ for $u\geq 1$ and $h'(u)\leq 0$ for $0<u\leq 1$.

We first show $h'(u)\leq 0$ for $0<u\leq 1$.
In $0<u\leq 1$, both
\eq{
\frac{u-1}{u(1+u)^2}\leq 0
}
and
\eq{
(1-c_0)u^2+(2-3c_0)u+1\geq 0-u+1\geq 0
}
hold due to $c_0<1$, which directly implies $h'(u)\leq 0$.

We next show $h'(u)\geq 0$ for $u\geq 1$.
In $u\geq 1$, both
\eq{
\frac{u-1}{u(1+u)^2}\geq 0
}
and
\eq{
(1-c_0)u^2+(2-3c_0)u+1=(1-c_0) \( u+\frac{2-3c_0}{2(1-c_0)}\) ^2 +\frac{c_0(8-9c_0)}{4(1-c_0)} 
\geq \frac{c_0(8-9c_0)}{4(1-c_0)} 
\geq 0
}
hold due to $c_0\leq 8/9$, which directly implies $h'(u)\geq 0$.

Combining them, we obtain the inequality \eqref{89ineq}, whose sum over all $i$ is equivalent to the desired inequality \eqref{rel-ent-lemma}:
\balign{
D(p||q)=&\sum_i p_i\ln \frac{p_i}{q_i}=\sum_i p_i\ln \frac{p_i}{q_i}+q_i-p_i \geq c_0 \sum_i \frac{(p_i-q_i)^2}{p_i+q_i}. \lb{lemma-pep}
}
}

\section{Proof of Lemma 2}\lb{s:lemma2}
\appnum{\Alph{section}}

We recast the Lemma 2 below:
\eqa{
(a-b)\ln \frac{a}{b}\geq \frac{2(a-b)^2}{a+b}.
}{app-lemma2}
In the following, we shall show the proof of this inequality.

\bpf{
Due to the symmetry, we set $a>b$ without loss of generality.
(In case of $a=b$, \eref{app-lemma2} is obviously satisfied.)
The inequality \eqref{app-lemma2} is equivalent to
\eq{
\ln a-\ln b\geq \frac{2(a-b)}{a+b}.
}
This relation directly follows from the downward-convexity of the function $1/x$:
\eq{
\ln a-\ln b=\int_b^a\frac{dx}{x}\geq (a-b)\frac{1}{\frac{a+b}{2}}=\frac{2(a-b)}{a+b}.
}
}

\section{Inequality on relative entropy}\lb{s:best-coeff}
\appnum{\Alph{section}}

We derived an inequality between relative entropy and triangular discrimination \eqref{rel-ent-lemma} in Sec.~\ref{s:Markov-gen}.
The obtained inequality is better than the existing one~\cite{Tan05}:
\eq{
D(p||q)\geq \frac{27}{32} \sum_i \frac{(p_i-q_i)^2}{p_i+q_i}.
}
However, our coefficient $c_0=8/9$ is still not the best one.
We here seek the best coefficient.

The crucial relation in the derivation is 
\eq{
a\ln \frac{a}{b}+b-a\geq c\frac{(a-b)^2}{a+b}.
}
We consider the maximum of $c$ satisfying the above inequality for any $a,b>0$.
As shown in \ref{s:lemma1}, this inequality is equivalent to 
\eq{
-\ln u+u-1-\frac{c(1-u)^2}{1+u}=:h(u)\geq 0
}
with $u>0$.
The local minimum of $h(u)$ for $c>8/9$ is calculated as
\eq{
u=1, \frac{3c-2+\sqrt{9c^2-8c}}{2(1-c)}.
}
We denote the second solution by $u^*(c)$.
Because $h(1)=0$, $h(u^*(c))\geq 0$ is the necessary and sufficient condition for $h(u)\geq 0$.
The relation $h(u^*(c))\geq 0$ is solved numerically as
\eq{
c\leq c^*:= 0.89612\cdots ,
}
whose right-hand side is the best coefficient for the inequality between relative entropy and triangular discrimination:
\eq{
D(p||q)\geq c^* \sum_i \frac{(p_i-q_i)^2}{p_i+q_i}.
}
We remark that the above coefficient is tightest because a nontrivial pair of probability distributions $p_1=1/(1+u^*(c^*))$, $p_2=u^*(c^*)/(1+u^*(c^*))$, $q_1=u^*(c^*)/(1+u^*(c^*))$, $q_2=1/(1+u^*(c^*))$ achieves its equality.

\section{Finiteness of $\Theta$}\lb{s:fin-theta}

In this Appendix, we show that $\Theta$ is finite under some physically-plausible assumptions.

\subsection{Upper bound of $\Theta^{(1)}$}\lb{s:fin-theta1}

We here derive some upper bounds of $\Theta^{(1)}$.
We first bound $\Theta_\mu^{(1)}$ defined in \eref{thetamu1} as
\balign{
\Theta_\mu^{(1)}(t)
:=&\frac{1}{c_0} \sum_{w\neq w'}(\Di E_w^{\mu, \lmd (t)})^2 (R_{ww'}^{\mu, \lmd (t)}p_{w',t}+R_{w'w}^{\mu, \lmd (t)}p_{w,t}) \nt \\
=&\frac{1}{c_0}\( \sum_{w,w'}(\Di E_w^{\mu, \lmd (t)})^2 R_{ww'}^{\mu, \lmd (t)}p_{w',t} -2\sum_w (\Di E_w^{i, \lmd (t)})^2R_{ww}^{\mu, \lmd (t)}p_{w,t} \) \nt \\
=&\frac{1}{c_0} \( \sum_{w}(\Di E_w^{\mu, \lmd (t)})^2 \[ \frac{d}{dt}\] _\mu p_{w,t}  -2\sum_w (\Di E_w^{\mu, \lmd (t)})^2R_{ww}^{\mu, \lmd (t)}p_{w,t} \) \nt \\ 
\leq&\frac{1}{c_0}\sum_{w^{-i}}p_{w^{-i}} \( \[ \frac{d}{dt}\]_\mu \la (\Di E_w^{\mu, \lmd (t)})^2\ra _{t,w^{-i}} +2R_{\rm max}\la (\Di E_w^{\mu, \lmd (t)})^2\ra_{t,w^{-i}}  \) .
}
Here, $R_{\rm max}$ is the maximum of the absolute value of the diagonal elements of the transition matrix (i.e., $|R_{ww}^{\mu, \lmd (t)}|\leq R_{\rm max}$ for all $w$ and $t$), and $\[ \frac{d}{dt}\]_\mu$ represents time derivative in case that the time evolution is induced only by $R^{\mu, \lmd (t)}$.
In the above calculation, we used the fact that $R^{\mu, \lmd (t)}$ keeps the distribution $p_{w^{-i}}$.
Because the fluctuation of the energy of the $i$-th particle $\langle (\Di E_w^{\mu, \lmd (t)})^2\rangle _{t,w^{-i}}$ and its time derivative are expected to be finite in physical systems, the above relation implies that $\Theta^{(1)}_\mu$ is finite if the diagonal elements of the transition matrix is bounded above.

In some cases, we can obtain the upper bound of $\Theta^{(1)}_\mu$ even though the diagonal elements of the transition matrix is unbounded.
To treat this situation, we again consider a system where only the particle $i$ is movable and other particles are fixed at $w^{-i}$.
We introduce a quantity $E_w^{i,\lmd (t)}$ which is the energy of the particle $i$ under the condition that other particles are fixed at $w^{-i}$.
Note that $E_w^{i,\lmd (t)}=O(1)$ with respect to the number of particles $M$.
We now state some requirements on the system.
First, we assume that the conditional probability distribution decays exponentially with energy: $p_{w^i,t|w^{-i}} \leq C_1\cdot e^{-a E_w^{i,\lmd (t)}}$ with constants $C_1$ and $a$.
This condition means that the probability distribution is not so far from a canonical distribution.
We also assume that both the diagonal term of the transition matrix and the number of states of the $i$-th particle below a certain energy $\Omega_{w^i|w^{-i}} (E)$ increase only polynomially with respect to energy: 
$|R_{ww}^{\mu, \lmd (t)}|\leq C_2 (E_w^{i,\lmd (t)})^b$ and $\Omega_{w^i|w^{-i}} (E)\leq C_3 E^c$ with constants $C_2$, $C_3$, $b$ and $c$, where $\mu =(i,\nu)$.
These conditions are expected to be satisfied in small systems with a finite number of particles including master-Boltzmann systems~\cite{Sie60, BKM04, FKS12}.
Under the aforementioned assumptions, $\Theta_\mu^{(1)}$ is bounded above as
\balign{
\Theta_\mu^{(1)} (t)
=&\frac{1}{c_0} \( \sum_{w}(\Di E_w^{\mu, \lmd (t)})^2 \[ \frac{d}{dt}\] _\mu p_{w,t}  -2\sum_w (\Di E_w^{\mu, \lmd (t)})^2R_{ww}^{\mu, \lmd (t)}p_{w,t} \) \nt \\ 
\leq&\frac{1}{c_0}\sum_{w^{-i}}p_{w^{-i}} \( \[ \frac{d}{dt}\]_\mu \la (\Di E_w^{\mu, \lmd (t)})^2\ra _{t,w^{-i}}  +2\sum_w (E_w^{\mu, \lmd (t)})^2R_{ww}^{\mu, \lmd (t)}p_{w^i,t|w^{-i}}   \) \nt \\
\leq&\frac{1}{c_0}\sum_{w^{-i}}p_{w^{-i}} \( \[ \frac{d}{dt}\]_\mu \la (\Di E_w^{\mu, \lmd (t)})^2\ra _{t,w^{-i}} +2C_1C_2C_3 \int_0^\infty dE E^{2+c+b}e^{-aE}   \) , \lb{bound-theta1}
}
where the integral $\int_0^\infty dx x^{2+c+b}e^{-ax}$ is obviously finite.

\subsection{Thermodynamic limit}\lb{s:thermo-theta}

The obtained inequalities \eqref{Jsigma-1} and \eqref{Jsigma-2} are still meaningful even in the thermodynamic limit.
In other words, the inequalities provide a nontrivial prediction for macroscopic systems.

The inequalities contain three terms, the heat current $\Jq$, the entropy production rate $\dsgm$, and the coefficient $\Theta$.
The former two terms are proportional to the system size $V$ or the particle number $M$ (More precisely, $\Jq_\nu$, $\hsgm$ and $\Theta$ are proportional to the volume of the region interacting with baths).
We first consider the case of $\Theta=\Theta^{(2)}$.
Because the number of states $w'$ satisfying $R_{w'w}\neq 0$ is proportional to $M$ with fixed $w=(w^1, \cdots ,w^M)$ and $(E_w-E_{w'})^2=O(1)$ is independent of $M$, $\Theta^{(2)}$ is also proportional to $M$.
We next confirm that $\Theta^{(1)}$ is proportional to $M$.
A single particle energy fluctuation $(\Di E_w^{\mu, \lmd (t)})^2$ defined in \eref{def-DiE} is independent of $M$, which leads to $\Theta^{(1)}_\mu=O(1)$.
Then, since $\Theta^{(1)}$ is the summation of it over $\mu$, $\Theta^{(1)}$ is proportional to $M$.
We remark that because $C_1=O(1/M)$, $C_2=O(1)$, and $C_3=O(M)$, the upper bound \eqref{bound-theta1} also scales in proportion to $M$.

We note that $\Theta^{(1)}$ is proportional to $M$ because we did not employ the energy fluctuation of the whole system $\Di E_w^{\lmd (t)}:=E_w^{\lmd (t)}-\la E\ra_t$ itself, but to decompose it into the contribution from each particle in the definition of $\Theta^{(1)}$.
In fact, if we define $\Theta '$ by using $\Di E_w^{\lmd (t)}$ as
\eq{
\Theta ':=\frac{1}{c_0}\sum_\mu \sum_{w\neq w'}(\Di E_w^{\lmd (t)})^2 (R_{ww'}^{\mu, \lmd (t)}p_{w',t}+R_{w'w}^{\mu, \lmd (t)}p_{w,t}),
}
then (although the inequality $\sum_\nu |\Jq_\nu| \leq \sqrt{\Theta' \dsgm}$ still holds) this coefficient $\Theta'$ is proportional to $M^2$.
This is because the energy fluctuation of the whole system $\Di E_w^{\lmd (t)}$ has variance of order $O(\sqrt{M})$.
In this case, the inequality in the thermodynamic limit gives no information more than the second law of thermodynamics.

\section{Extension of \eref{GJsigma-2} to the case of finite time interval}\lb{s:extend-tur}

Using the techniques that we have introduced, we can derive a similar but still different relation to the thermodynamic uncertainty relation.
To this end, we consider a process in a finite-time interval $0\leq t\leq \tau$ in stationary state with the local detailed-balance condition.
Owing to the local detailed-balance condition, the entropy production rate is written as
\eq{
\dsgm =\sum_{w,w'}R_{ww'}p_{w'} \ln \frac{R_{ww'}p_{w'}}{R_{w'w}p_w}.
}
We denote a single trajectory of time evolution in $0\leq t\leq \tau$ by $\Gamma$ and its time-reversal by $\Gamma^\dagger$.
We also denote the probability density for the realization of $\Gamma$ by $P(\Gamma)$.
The average of a stochastic variable is denoted by $\la \cdot \ra$.
It is well known that the total entropy production $\Sigma:=\int_0^\tau dt \dsgm(t) =\tau \dsgm$ is written as~\cite{Sei12}
\eq{
\Sigma =\int d\Gamma P(\Gamma)\ln \frac{P(\Gamma)}{P(\Gd)}.
}

Let $X(\Gamma)$ be a time-asymmetric stochastic variable (i.e., $X(\Gamma)=-X(\Gamma^\dagger)$), which includes any current of a conserved quantity.
Then, $X$ satisfies the following theorem:

\

{\bf Theorem 5}:
In a Markov process with finite time interval $\tau$, we have
\eqa{
\la X^2\ra \Sigma \geq 2\la X \ra ^2.
}{ftur}
Here, we normalize the Boltzmann constant to unity.

\

\bpf{
We employ the same technique as the derivation of \eref{Jsigma-2}.
With noting $\int d\Gamma X(\Gamma)P(\Gamma)=-\int d\Gamma X(\Gd)P(\Gamma)$ due to the time-asymmetric property of $X$, we transform $\abs{\la X\ra}^2$ as
\balign{
\abs{ \la X\ra}^2
&=\abs{\int d\Gamma X(\Gamma)P(\Gamma )}^2 \nt \\
&=\abs{\frac{1}{2}\int d\Gamma X(\Gamma)(P(\Gamma )-P(\Gd))}^2 \nt \\
&=\abs{\frac{1}{2}\int d\Gamma X(\Gamma)\sqrt{P(\Gamma)+P(\Gd)} \cdot \frac{P(\Gamma )-P(\Gd)}{\sqrt{P(\Gamma)+P(\Gd)} }}^2 \nt \\
&\leq \frac{1}{4}\( \int d\Gamma X(\Gamma)^2 (P(\Gamma)+P(\Gd))\) \cdot \( \int d\Gamma \frac{(P(\Gamma )-P(\Gd))^2}{P(\Gamma)+P(\Gd)} \) . \lb{mid-1}
}

In a similar manner to \eref{Jq-db}, we have the following expression of the entropy production:
\balign{
\Sigma &=\int d\Gamma P(\Gamma)\ln \frac{P(\Gamma)}{P(\Gd)}=\frac{1}{2}\int d\Gamma (P(\Gamma)-P(\Gd)) \ln \frac{P(\Gamma)}{P(\Gd)} \geq \int d\Gamma \frac{(P(\Gamma)-P(\Gd))^2}{P(\Gamma)+P(\Gd)}.
}
Finally, inserting the following relation
\eq{
\int d\Gamma X(\Gamma)^2 (P(\Gamma)+P(\Gd)) 
=2 \int d\Gamma (X(\Gamma))^2P(\Gamma)
=2\la X^2\ra
}
into \eref{mid-1}, we arrive at the desired inequality:
\balign{
\abs{ \la X\ra}^2
&\leq \frac{1}{4}\( \int d\Gamma X(\Gamma)^2 (P(\Gamma)+P(\Gd))\) \cdot \( \int d\Gamma \frac{(P(\Gamma )-P(\Gd))^2}{P(\Gamma)+P(\Gd)} \) \nt \\
&\leq \frac12 \la X^2\ra \Sigma .
}
}

\section{Efficiency of thermoelectric transport}\lb{s:eff-te}
\appnum{\Alph{section}}

We here briefly see how to define efficiency in a stationary thermoelectric transport system considered in \sref{thermoele}.
Same as \sref{thermoele}, we consider two heat-particle baths with inverse temperatures and chemical potentials $\beta_1, \mu_1$ and $\beta_2, \mu _2$, respectively.
We set $\beta_1<\beta_2$, $\mu_1<\mu_2$.
The heat and particle currents from the bath 1 to 2 are denoted by $\Jq$ and $\Jn$, both of which we assume positive.
Namely, the particle current $\Jn$ flows against chemical potential gradient, which we regard as {\it work}.

In a cyclic process, efficiency is defined as $W/\QH$ with $\QH$ as heat absorption from the hot bath.
We now define the counterpart of $\QH$ in thermoelectric transport.
Because particles themselves have their own energy in the form of chemical potential, we subtract this from heat current and regard $\Jq -\mu_1\Jn$ as the counterpart of $\QH$.
Thus, we define efficiency in thermoelectric transport as
\eqa{
\eta :=\frac{\Di \mu \Jn}{\Jq -\mu_1\Jn},
}{te-eta}
where we defined $\Di \mu :=\mu_2-\mu_1>0$ and assumed $\Jq -\mu_1\Jn>0$.

We now confirm that the efficiency is indeed bounded by the Carnot efficiency
\eq{
\etac :=1-\frac{\beta_1}{\beta_2}\geq \eta .
}
The above inequality is equivalent to
\eq{
(\beta_2-\beta_1)(\Jq-\mu_1\Jn )\geq \beta_2 (\mu_2-\mu_1)\Jn
}
which is transformed into the nonnegativity of entropy production rate
\eq{
\dot{\sigma}=(\beta_2-\beta_1)\Jq +(\beta_1\mu_1-\beta_2\mu_2)\Jn \geq 0.
}

\section{Failure of finite-time thermodynamic uncertainty relation in relaxation process}\lb{s:relax-tur}
\appnum{\Alph{section}}

We show that the thermodynamic uncertainty relation $\la \Di X^2\ra \Sigma \geq 2\la X \ra ^2$ holds only in stationary system, and cannot be extended to relaxation processes with time-independent transition matrix satisfying local detailed-balance condition.

Consider a stochastic process on two states $w\in \{ 1,2\}$ with the same energy.
The transition matrix thus satisfies $R_{12}=R_{21}$.
We set $X$ as time integration of probability current from 1 to 2.
Suppose that the initial distribution at $t=0$ is $p_1(0)=1$ and $p_2(0)=0$, and consider the long time limit $t\to \infty$, where the distribution relaxes to equilibrium distribution $p_1(\infty)=p_2(\infty)=1/2$.

Straightforward calculation tells
\balign{
\la X\ra &=\frac{1}{2} \\
\la \Di X^2\ra &=\la X^2\ra -\la X\ra ^2 =\frac{1}{4} \\
\Sigma &= \ln 2 - 0 = \ln 2 (=0.6921\cdots ).
}
Hence, $\la \Di X^2\ra \Sigma =(\ln 2)/4< 1/4$ and $2\la X \ra ^2 =1/2$, which obviously violates the extended thermodynamic uncertainty relation in relaxation processes.

\end{document}